\documentclass[twocolumn,english,prl,floatfix,citeautoscript,nofootinbib]{revtex4}
\usepackage{amsbsy}
\usepackage{latexsym,epsfig,graphicx}
\usepackage{dcolumn}
\usepackage{subfigure}
\usepackage{comment}
\usepackage{color}
\usepackage[colorlinks,urlcolor=blue,citecolor=blue]{hyperref}
\usepackage{amstext}
\usepackage{amssymb}
\usepackage{setspace}

\begin{document}

\title{Higher-order topological corner states induced by gain and loss}
\author{Xi-Wang Luo}
\author{Chuanwei Zhang}
\thanks{Corresponding author. \\
Email: \href{mailto:chuanwei.zhang@utdallas.edu}{chuanwei.zhang@utdallas.edu}%
}
\affiliation{Department of Physics, The University of Texas at Dallas, Richardson, Texas
75080-3021, USA}

\begin{abstract}
Higher-order topological insulators and superconductors are topological
phases that exhibit novel boundary states on corners or hinges. Recent
experimental advances in controlling dissipation such as gain/loss in atomic
and optical systems provide a powerful tool for exploring non-Hermitian
topological phases. Here we show that higher-order topological corner states
can emerge by introducing staggered on-site gain/loss to a Hermitian system
in a trivial phase. For such a non-Hermitian system, we establish a general
bulk-corner correspondence by developing a biorthogonal nested-Wilson-loop
and edge-polarization theory, which can be applied to a wide class of
non-Hermitian systems with higher-order topological orders. The theory gives
rise to topological invariants characterizing the non-Hermitian topological
multipole moments (i.e., corner states) that are protected by reflection or
chiral symmetry. Such gain/loss induced higher-order topological corner
states can be experimentally realized using photons in coupled cavities or
cold atoms in optical lattices.
\end{abstract}

\maketitle

\emph{\textcolor{blue}{Introduction}.---} Topological states of matter \cite%
{xiao2010berry,hasan2010colloquium,qi2011topological,RevModPhys.88.035005,Sato2017Topological}
have been widely studied in various systems ranging from solid-state~\cite%
{PhysRevLett.95.146802,Bernevig2006quantum,Konig2007quantum}, over cold
atomic~\cite{PhysRevLett.111.185301,PhysRevLett.111.185302,Goldman2014light,
Aidelsburger2014measuring,Jotzu2014Experimental,Li2016Bloch,Flaschner2016Experimental, Goldman2016topological,Cooper2018topological}
to photonic~\cite%
{haldane2008possible,hafezi2011robust,fang2012realizing,lu2014topological,kraus2012topological,Hafezi2013imaging,Ozawa2018Topological}
and acoustic~\cite%
{He2018Topological,Xiao2015Synthetic,PhysRevLett.120.116802,PhysRevLett.114.114301,He2016Acoustic,Ma2019Topological}
systems. The states are indexed by the bulk topological invariants that
determine the boundary physics with lower dimensions. Recently, the concept
has been generalized to higher-order topological insulators or
superconductors with novel boundary states on corners or hinges~\cite%
{PhysRevLett.110.046404,Benalcazar2017Quantized,PhysRevLett.119.246401,PhysRevLett.119.246402,PhysRevB.96.245115,PhysRevB.98.241103,Schindler2018Higher,Imhof2018Topolectrical-circuit,Hassan2018Corner,Serra-Garcia2018observation,arXiv1812.09304,Peterson2018a,PhysRevLett.120.026801,PhysRevB.97.241405,PhysRevB.97.205135,PhysRevLett.121.186801,PhysRevLett.121.096803,PhysRevX.9.011012,PhysRevB.98.205147,PhysRevLett.121.196801,arXiv1901.07579}%
. Different from conventional first-order topological states, the $d$%
-dimensional $n$-th order topological states can host $(d-n)$-dimensional
gapless boundary states. The experimental realizations of such interesting
higher-order topological states in photonic~\cite%
{Hassan2018Corner,Serra-Garcia2018observation,PhysRevB.98.205147,arXiv1812.09304}
and electrical circuit~\cite{Imhof2018Topolectrical-circuit,Peterson2018a}
systems further enlighten the research of these novel topological matters.

Meanwhile, the search for topological states of matter has also turned to
open quantum systems characterized by non-Hermitian Hamiltonians~\cite%
{Rept.Prog.Phys.70.947}, which exhibit a rich variety of unique properties
without Hermitian counterparts~\cite{Eur.Phys.J.Spec.Top227}. States modeled
by non-Hermitian Hamiltonians appear in systems such as photonic structures
with loss or gain~\cite{PhysRevLett.100.103904, PhysRevLett.115.200402,
PhysRevLett.115.040402,NatMater16433,Photon.Rev.6.A51,
Nature488.167,PhysRevLett.113.053604,Science346Loss,Zhao2018topological,PhysRevLett.120.113901,Bandres2018Topological,St-Jean2017Lasing}%
, and cold atomic systems or solid-state materials with finite
(quasi-)particle lifetime~\cite%
{Ashida2017Parity,PhysRevLett.121.026403,arXiv1802.00443,PhysRevB.98.035141,PhysRevLett.118.045701,arXiv1608.05061,Muller2012Engineered,arXiv1811.06046}%
. The eigenvalues are generally complex, and the right and left eigenstates,
satisfying biorthonormolity constrains, are no longer equivalent to each
other (neither of them form an orthogonal basis).
Moreover, more than one right eigenstates can coalesce at exceptional points~%
\cite{PhysRevLett.118.045701}. Such unique properties lead to a rich variety
of interesting topological phenomena (e.g., the non-Hermitian skin effects,
exceptional rings, bulk fermi arcs, etc.), with bulk-boundary correspondence
very different from the Hermitian systems~\cite%
{PhysRevB.84.205128,PhysRevLett.116.133903, PhysRevLett.121.136802,
PhysRevLett.121.213902,
PhysRevLett.121.086803,PhysRevX.8.031079,PhysRevLett.121.026808,arXiv1808.09541,PhysRevB.97.045106, PhysRevLett.118.040401,PhysRevLett.120.146402,PhysRevA.93.062101,Xiong2018why,Science359Observation, arXiv1804.04676,PhysRevB.99.081103,arXiv1812.02011,arXiv1809.02125,arXiv1902.07217}%
.

The effects of non-Hermiticity on higher-order topological physics have been
considered recently in a few works ~\cite%
{arXiv1810.04067,arXiv1811.12059,arXiv1810.04527,arXiv1810.11824,arXiv1812.09060}%
, where the non-Hermiticity is induced by asymmetric tunnelings, leading to
the observation of interesting phenomena such as higher-order skin effect~%
\cite{arXiv1810.04067} and biorthogonal bulk polarization~\cite%
{arXiv1812.09060}. Nevertheless, a general bulk-corner correspondence of the
non-Hermitian higher-order topological states is still elusive. In addition,
compared to asymmetric tunnelings, a simpler and more tunable way for
introducing non-Hermiticity in photonic and atomic experiments is to control
the on-site particle dissipations directly. Therefore two natural questions
arise: \textit{i}) Can higher-order topological states be induced by simply
controlling the on-site gain or loss? \textit{ii}) Is there a general
bulk-corner correspondence for the non-Hermitian higher-order topological
states?

In this Letter, we address these two important questions by considering a
2-dimensional (2D) lattice model with staggered on-site particle gain/loss.
Our main results are:

\textit{i}) The non-Hermitian particle gain and loss can drive the system
from a trivial phase to a second-order topological phase with the emergence
of four degenerate corner states.

\textit{ii}) We develop the biorthogonal nested-Wilson-loop and
edge-polarization approach which gives rise to bulk topological invariants
responsible for the gapless corner states. The topological invariants
are protected by reflection or chiral symmetries.
In the presence of additional $C_{4}$ rotation symmetry, the topology can
also be characterized by a quantized biorthogonal winding number.

\textit{iii}) Although we focus on 2D reflection-symmetric case, our model
and the bulk-corner correspondence
can be generalized to study $d$-dimensional $d$-th order non-Hermitian
topological states with either reflection or chiral symmetries.

\textit{iv}) Simple experimental schemes based on photons in coupled
cavities and cold atoms in optical lattices are proposed. Our system only
relies on the manipulation of on-site particle gain/loss, and is ready for
experimental exploration.

\begin{figure}[t]
\includegraphics[width=1.0\linewidth]{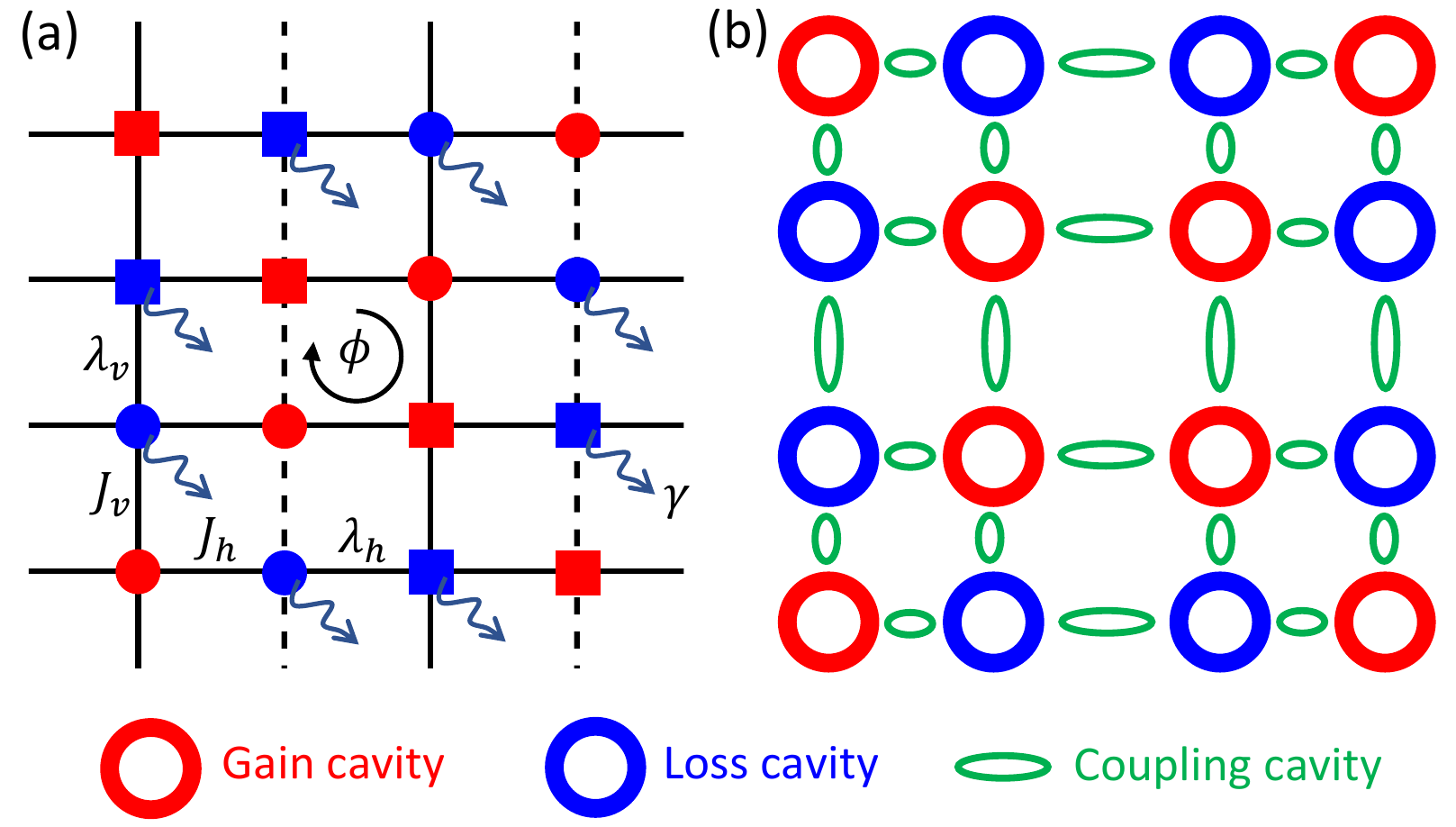}
\caption{(a) Lattice representation of the non-Hermitian model in Eq.~%
\protect\ref{Eq:Ham}. All sites in blue (red) have particle loss (gain) with
a rate $\protect\gamma$. $\protect\phi$ is the magnetic flux for each
plaquette, and $J_{h,v}$ ($\protect\lambda_{h,v}$) are the tunneling
amplitudes between sites in different color (shape) along the horizontal and
vertical directions, respectively. (b) Experimental implementation of the
lattice model in (a) using coupled arrays of micro-ring cavities.}
\label{fig:sys}
\end{figure}

\emph{\textcolor{blue}{The model}.---} We consider a 2D lattice model with
staggered tunnelings along both horizontal and vertical directions,
as shown in Fig.~\ref{fig:sys}(a). There is an effective magnetic flux $\phi
=\pi $ 
for each plaquette, which appears as the tunneling phases on the dashed
lines. The non-Hermiticity is introduced by the particle loss (gain) on all
blue (red) lattice sites. We choose 16 orbitals in Fig.~\ref{fig:sys}(a) as
our unit cell with horizontal and vertical primitive-lattice vectors. The
Hamiltonian reads 
\begin{eqnarray}
H(\mathbf{k}) &=&J_{h}\sigma _{h}^{x}+J_{v}\sigma _{v}^{x}\sigma _{h}^{\phi
}+i\gamma \sigma _{h}^{z}\sigma _{v}^{z}\tau _{h}^{z}\tau _{v}^{z}  \nonumber
\\
&&+\lambda _{h}(\tau _{h}^{-}\sigma _{h}^{+}+e^{-ik_{x}}\tau _{h}^{-}\sigma
_{h}^{-}+h.c.)  \label{Eq:Ham} \\
&&+\lambda _{v}\sigma _{h}^{\phi }(\tau _{v}^{-}\sigma
_{v}^{+}+e^{-ik_{y}}\tau _{v}^{-}\sigma _{v}^{-}+h.c.),  \nonumber
\end{eqnarray}%
where $J_{h,v}>0$ ($\lambda _{h,v}>0$) are the nearest-neighbour tunneling
amplitudes between red and blue (circle and square) sites, $\mathbf{\sigma }%
_{h,v}$ ($\mathbf{\tau }_{h,v}$) are the Pauli matrices for the degrees of
freedom spanned by red and blue (circle and square) sites, and $h,v$
represent the horizontal and vertical directions, respectively. $\sigma
_{h,v}^{\phi }=\sigma _{h,v}^{z}$ for $\phi =\pi $. The gain/loss rate $%
\gamma $ in Eq.~\ref{Eq:Ham} is positive since the blue sites are lossy.
Alternatively, we may consider a different gain/loss configuration with gain
(loss) on blue (red) sites, which simply changes $\gamma $ to negative. In
experiments, the Hamiltonian can be realized using cold atoms in optical
lattices or photons in coupled cavities~\cite{SM}. Fig.~\ref{fig:sys} (b) is
an example based on arrays of coupled micro-ring cavities, where the
coupling amplitude and phase between neighbour cavities, and the photon
gain/loss for each cavity can be controlled independently~\cite%
{arXiv1812.09304,Zhao2018topological}.



\begin{figure}[t]
\includegraphics[width=1.0\linewidth]{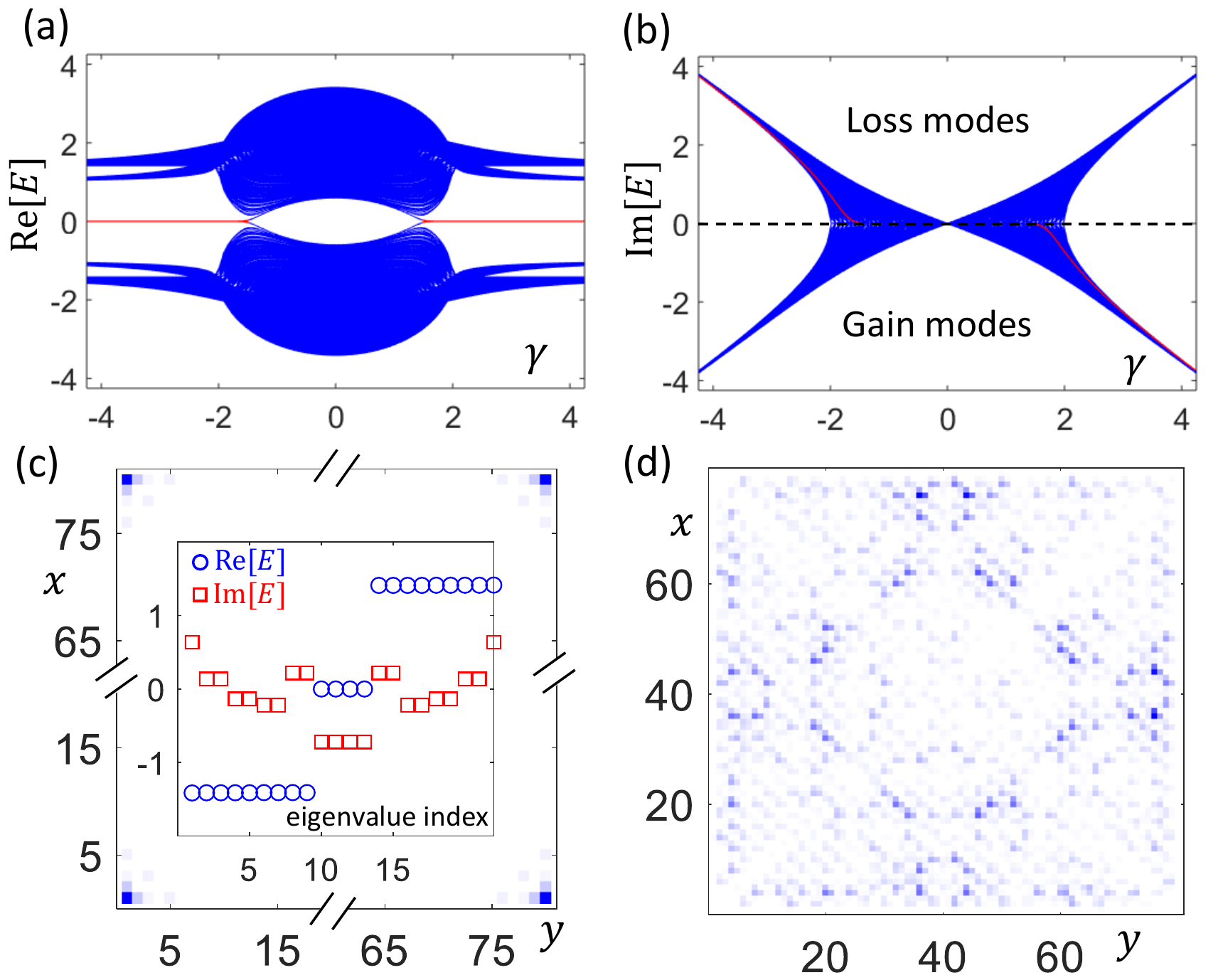}
\caption{(a) (b) Energy spectra of the non-Hermitian Hamiltonian Eq.~\protect
\ref{Eq:Ham} with open boundaries in both directions. The bulk energy gap
closes at $|\protect\gamma |=\protect\gamma _{c}$, where a topological phase
transition occurs and in-gap corner states (red curves with four-fold
degeneracy) emerge at $|\protect\gamma |>\protect\gamma _{c}$. (c) Typical
density distributions $|\Psi _{\text{corner}}^{R}(x,y)|^{2}$ of the four
corner states, with $|\Psi _{\text{corner}}^{R}(x,y)\rangle $ the right
eigenstate. The inset shows the corresponding eigenenergies around the four
corner states with $\text{Re}[E]=0$. (d) Typical density distributions of
the bulk states. $\protect\gamma =2$ in (c) and (d). Common parameters:
system size $N_{h}=N_{v}=20$ (unit cells), $J=\protect\sqrt{2}$ and $\protect%
\lambda _{v}=1$ (leading to $\protect\gamma _{c}=\protect\sqrt{2}$). We set $%
\protect\lambda _{h}=1$ as energy unit.}
\label{fig:corner}
\end{figure}

\emph{\textcolor{blue}{Corner states}.---} For simplicity, we assume $%
J_{h}=J_{v}\equiv J$ throughout this paper, and the physics for $J_{h}\neq
J_{v}$ is similar. The system has 16 bands~\cite{SM}, which appear in pairs $%
E(\mathbf{k})=-E^{\ast }(\mathbf{k})$ due to the pseudo-anti-Hermiticity $%
\eta H\eta =-H^{\dagger }$ with $\eta =\sigma _{h}^{z}\sigma _{v}^{z}\tau
_{h}^{z}\tau _{v}^{z}$. We are interested in the half-filling gap around $%
\text{Re}[E]=0$. We focus on the region $\lambda _{h(v)}\leq J$ (the system
stays in the trivial insulating/metal phase at the Hermitian limit $\gamma
=0 $~\cite{Benalcazar2017Quantized,PhysRevB.96.245115}), and show that the
second-order topological corner states can be induced solely by
non-Hermitian gain and loss.

In Figs.~\ref{fig:corner}(a) and (b), we plot the energy spectrum as a
function of $\gamma $, with open boundaries along both directions.
Effectively, the particle loss reduces the tunnelings between gain and loss
sites, while the tunnelings between two loss (gain) sites are not affected.
We see that as $|\gamma |$ increases, the bulk gap closes and reopens (the
small derivation is the finite size effect) at a critical point $\gamma _{c}$%
, leading to a topological phase transition with the emergency of four
in-gap states. The typical density distributions of these in-gap states are
shown in Fig.~\ref{fig:corner}(c), which are well localized at four corners.
We emphasize that our system does not suffer from the non-Hermitian skin
effects due to the trivial eigenenergy vorticity~\cite{PhysRevX.8.031079} $%
\oint \partial _{\mathbf{k}}\text{Arg}[E(\mathbf{k})]d\mathbf{k}=0$ for any
loop in the momentum space, therefore it does not matter whether the right
or/and left eigenstates are used to calculate the density distribution. As a
result, the bulk states of $H(\mathbf{k})$ do distribute in the bulk [see
Fig.~\ref{fig:corner}(d)], and the open-boundary bulk spectrum is the same
as that for periodic boundaries.
We set $\lambda _{h}=\lambda _{v}$ in Fig.~\ref{fig:corner}, therefore the
system undergoes a bulk gap closing across the topological phase transition
due to the $C_{4}$ symmetry~\cite{PhysRevB.97.205135}. In general, the
second-order topology can be altered by the gap closing in either the bulk
or edge spectrum, and the emergency of corner states does not require bulk
energy gap closing for $\lambda _{h}\neq \lambda _{v}$~\cite%
{PhysRevB.96.245115,PhysRevB.97.205135}, which will be further illustrated.

\emph{\textcolor{blue}{Topological invariants}.---} 
For Hermitian systems, it was shown that the topology of the nested Wilson
loop and edge polarization are responsible for the corner states~\cite%
{Benalcazar2017Quantized, PhysRevB.96.245115}. Here we develop their
non-Hermitian counterparts and show that the non-Hermitian corner states are
originated from the topology of the generalized biorthogonal nested Wilson
loops and edge polarizations. We consider a general Hamiltonian $H(\mathbf{k}%
)$ on a torus with periodic boundaries and define the biorthogonal Wilson
loop operator as
\begin{equation}
W_{h,\mathbf{k}}=\mathcal{P}\exp [i\int_{k_{x}}^{k_{x}+2\pi
}A_{h}(k_{x}^{\prime },k_{y})dk_{x}^{\prime }]\text{,}  \label{Eq:Wilson}
\end{equation}%
where $A_{h}(\mathbf{k})=-i\langle u_{n,\mathbf{k}}^{L}|\partial
_{k_{x}}|u_{m,\mathbf{k}}^{R}\rangle $ is the biorthogonal non-Abelian Berry
connection in the horizontal direction, $|u_{m,\mathbf{k}}^{R,L}\rangle $
are the $m$-th occupied right and left Bloch eigenstates satisfying $H(%
\mathbf{k})|u_{m,\mathbf{k}}^{R}\rangle =E_{m}(\mathbf{k})|u_{m,\mathbf{k}%
}^{R}\rangle $, $H^{\dagger }(\mathbf{k})|u_{m,\mathbf{k}}^{L}\rangle
=E_{m}^{\ast }(\mathbf{k})|u_{m,\mathbf{k}}^{L}\rangle $ and $\langle u_{n,%
\mathbf{k}}^{L}|u_{m,\mathbf{k}}^{R}\rangle =\delta _{n,m}$, and $\mathcal{P}
$ is the path-ordering operator. Different from the Hermitian case~\cite%
{Benalcazar2017Quantized}, $W_{h,\mathbf{k}}$ may no longer be a unitary
operator, and leads to a non-Hermitian Wannier Hamiltonian $H_{W_{h}}(%
\mathbf{k})=-\frac{i}{2\pi }\log W_{h,\mathbf{k}}$,
which also has different left and right eigenstates, that is, $H_{W_{h}}(%
\mathbf{k})|\varepsilon _{h,j,\mathbf{k}}^{R}\rangle =\varepsilon
_{h,j,k_{y}}|\varepsilon _{h,j,\mathbf{k}}^{R}\rangle $, $H_{W_{h}}^{\dagger
}(\mathbf{k})|\varepsilon _{h,j,\mathbf{k}}^{L}\rangle =\varepsilon
_{h,j,k_{y}}^{\ast }|\varepsilon _{h,j,\mathbf{k}}^{L}\rangle $ with $%
\langle \varepsilon _{h,j,\mathbf{k}}^{L}|\varepsilon _{h,j^{\prime },%
\mathbf{k}}^{R}\rangle =\delta _{j,j^{\prime }}$ and $j$ the Wannier band
index. The non-Hermitian 
Wannier bands (independent from $k_{x}$), which obey the identification $%
\text{Re}[\varepsilon _{h,j,k_{y}}]\equiv \text{Re}[\varepsilon _{h,j,k_{y}}]%
\text{ mod }1$, can carry topological invariants if they are gapped.

The biorthogonal vertical polarization for the Wannier band sector $%
\varepsilon _{h}$ can be defined as
\begin{equation}
p_{v}^{\varepsilon _{h}}=-\frac{i}{4\pi ^{2}}\int dk_{x}\log \det [\tilde{W}%
_{h,\mathbf{k}}]\text{.}  \label{Eq:Top_inv}
\end{equation}%
Here $\tilde{W}_{h,\mathbf{k}}$ is the biorthogonal nested Wilson loop along
the vertical direction, which is defined on the Wannier sector $\varepsilon
_{h}$ with non-Hermitian Wannier-band basis $|w_{h,j,\mathbf{k}%
}^{R(L)}\rangle =\sum_{m=1}^{N_{\text{occ}}}|u_{m,\mathbf{k}}^{R(L)}\rangle
\lbrack |\varepsilon _{h,j,\mathbf{k}}^{R(L)}\rangle ]_{m}$ ($N_{\text{occ}}$
is the number of occupied energy bands and $\langle w_{h,j,\mathbf{k}%
}^{L}|w_{h,j^{\prime },\mathbf{k}}^{R}\rangle =\delta _{j,j^{\prime }}$)~%
\cite{SM}. Similarly, we can obtain the biorthogonal nested Wilson loop
along the horizontal direction and the corresponding polarization $%
p_{h}^{\varepsilon _{v}}$. There would be corner states when $%
p_{h,v}^{\varepsilon _{v,h}}$ are non-trivial.

On the other hand, even for trivial $p_{h,v}^{\varepsilon _{v,h}}$, one may
still have corner states if the edge polarization is non-trivial~\cite%
{PhysRevB.96.245115}. For non-Hermitian systems, we should use the
biorthogonal edge polarization,
which are obtained by considering a cylindrical geometry and calculating the
pseudo-one-dimensional biorthogonal Wannier values ($\varepsilon _{h,j}$ or $%
\varepsilon _{v,j}$) and polarization ($p_{h}^{i_{v}}$ or $p_{v}^{i_{h}}$
with $i_{v}$ or $i_{h}$ the unit-cell index along the open direction) along
the periodic direction (horizontal or vertical)~\cite{SM}.
The second-order corner modes are characterized by the vanishing bulk
polarization (\textit{i.e.}, $i_{v,h}$ away from 1 and $N_{v,h}$), but
quantized non-zero edge-localized polarization $p_{h}^{\text{edge}}$ and/or $%
p_{v}^{\text{edge}}$ (\textit{i.e.}, $i_{v,h}$ near $1$ or $N_{v,h}$)~\cite%
{SM}.

In general, higher-order topological phases are protected by symmetries~\cite%
{Benalcazar2017Quantized, PhysRevB.96.245115}. We consider a Hamiltonian
that respects either reflection symmetries $%
M_{h}H(k_{x},k_{y})M_{h}^{-1}=H(-k_{x},k_{y})$ and $%
M_{v}H(k_{x},k_{y})M_{v}^{-1}=H(k_{x},-k_{y})$, or chiral (sublattice)
symmetry $\Xi H(k_{x},k_{y})\Xi ^{-1}=-H(k_{x},k_{y})$, with symmetry
operators given by $M_{h}$, $M_{v}$ or $\Xi $.
Since the biorthogonal Wannier bands or values (on a torus or cylinder)
change the signs under reflection operation, they are either flat bands
locked at $0$ or $\frac{1}{2}$, or appear in $\pm \varepsilon $ pairs for
reflection-symmetric systems. The reflection symmetries also ensure the
quantization of ($p_{h}^{\varepsilon _{v}}$, $p_{v}^{\varepsilon _{h}}$) and
($p_{h}^{\text{edge}}$, $p_{v}^{\text{edge}}$) with value $0$ or $\frac{1}{2}
$. Similar properties hold for the chiral-symmetric systems with
non-Hermiticity induced by asymmetric tunneling~\cite{SM}.

The Wannier bands correspond to the position of the particle density cloud~%
\cite{Benalcazar2017Quantized,PhysRevB.96.245115}. We focus on the Wannier
sectors $\in (0,\frac{1}{2})$ [or $\in (\frac{1}{2},1)$] which are
responsible for the edge topology and corner states. Based on the
Wannier-sector and edge polarizations, we define two topological invariants:
$Q_{1}=4p_{v}^{\varepsilon _{h}}p_{h}^{\varepsilon _{v}}$ mod 2 [with $%
\varepsilon _{v,h}$ the Wannier sector $\in (0,\frac{1}{2})$] and $%
Q_{2}=2(p_{h}^{\text{edge}}+p_{v}^{\text{edge}})$ mod 2. For the topological
phase, we have either $Q_{1}=1$ or $Q_{2}=1$; while for the trivial phase,
we have both $Q_{1}=0$ and $Q_{2}=0$. The above bulk-corner correspondence
can apply to any non-Hermitian systems with reflection or chiral symmetries,
and are reduced to the normal nested-Wilson-loop and edge-polarization
theory~\cite{Benalcazar2017Quantized} in the Hermitian limit.

\begin{figure}[t]
\includegraphics[width=1.0\linewidth]{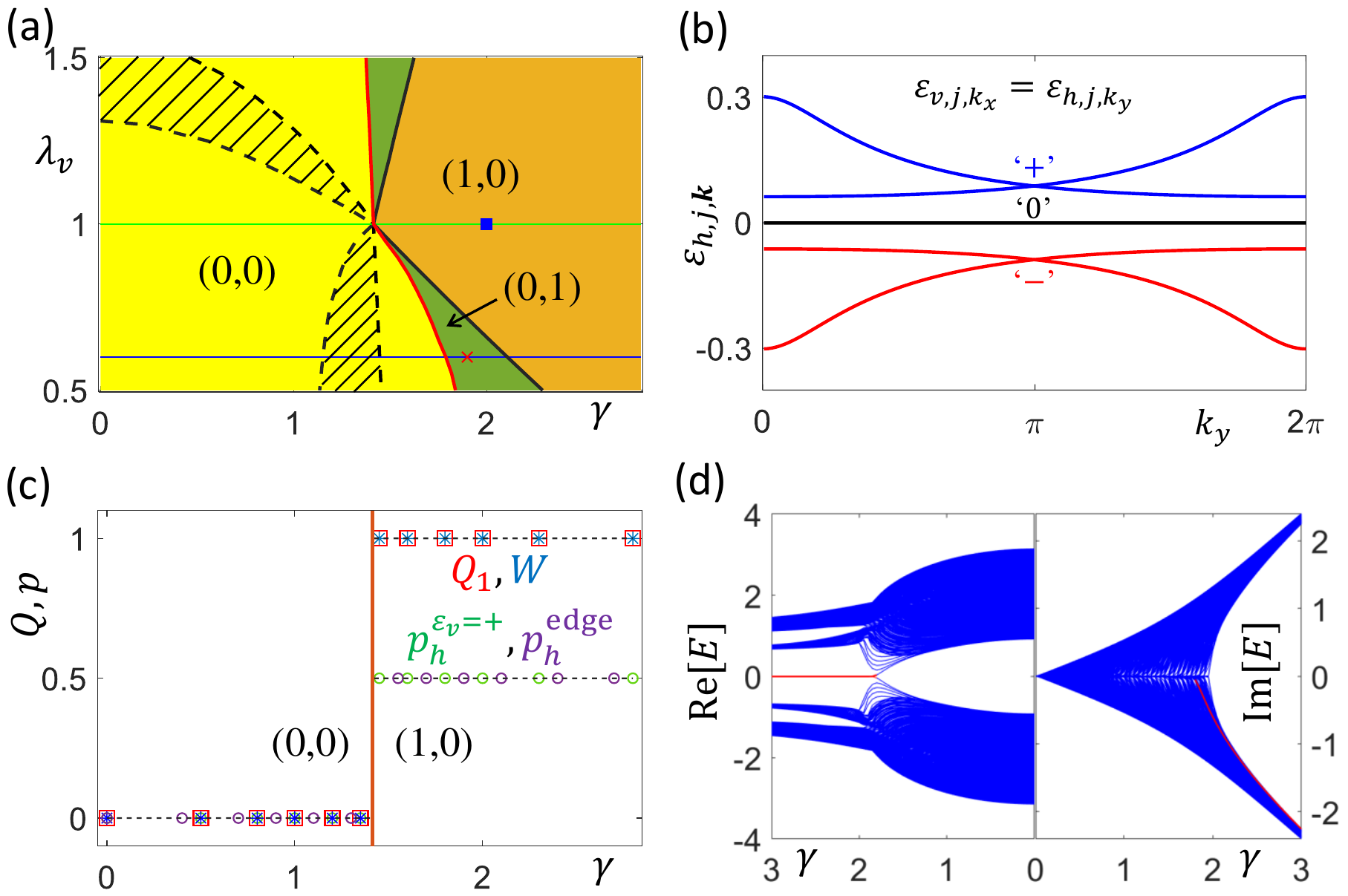}
\caption{(a) Phase diagram in the $\protect\gamma $-$\protect\lambda _{v}$
plane for $J=\protect\sqrt{2}\protect\lambda _{h}$, with one trivial phase
[yellow area with $(Q_{1},Q_{2})=(0,0)$] and two topological phases [T-I:
orange area with $(1,0)$ and T-II: green area with $(0,1)$]. The patterned
region has a vanishing Wannier-band gap~\protect\cite{SM}. (b) Wannier band
structures with $\protect\lambda _{v}=\protect\lambda _{h}$ and $\protect%
\gamma =2$ [the blue square in (a)]. The imaginary parts are locked at 0.
(c) Wannier-sector (green circles) and edge (purple circles) polarizations
as well as topological invariant $Q_{1}$ (red squares) and winding number $W$
(blue stars), with $\protect\lambda _{v}=\protect\lambda _{h}$ [the thin
green line in (a)]. (d) Complex energy spectra with open boundaries [$%
N_{h}=N_{v}=20$ (unit cells)] and $\protect\lambda _{v}=0.6\protect\lambda %
_{h}$ [the thin blue line in (a)]. The bulk gap persists upon the phase
transition. We set $\protect\lambda _{h}=1$ as the energy unit.}
\label{fig:phase}
\end{figure}

\emph{\textcolor{blue}{Phase diagram}.---} As an example, we study the phase
diagram of the model in Fig.~\ref{fig:sys} based on the biorthogonal
topological invariants. The corresponding Hamiltonian satisfies reflection
symmetries with $M_{h}=\sigma _{v}^{\phi }\sigma _{h}^{x}\tau _{h}^{x}$ and $%
M_{v}=\sigma _{v}^{x}\tau _{v}^{x}$. It also possesses the rotational
symmetry $C_{4}H(k_{x},k_{y})C_{4}^{-1}=H(k_{y},-k_{x})$ if $\lambda
_{h}=\lambda _{v}$, where $C_{4}=C_{\tau }\otimes C_{\sigma }$ with
\begin{equation}
C_{\tau }=\frac{1}{2}\sum_{s\neq \bar{s}}\left[ \tau _{s}^{+}(\frac{1-\tau _{%
\bar{s}}^{z}}{2})+\tau _{s}^{-}(\frac{1+\tau _{\bar{s}}^{z}}{2})\right] ,
\end{equation}%
and $C_{\sigma }$ has a similar expression with $s,\bar{s}=\{h,v\}$. In Fig.~%
\ref{fig:phase}(a), we show the phase diagram in the $\gamma $-$\lambda _{v}$
plane with $J/\lambda _{h}=\sqrt{2}$. The phase diagram is symmetric with
respect to $\gamma =0$, so we focus on $\gamma \geq 0$. The right and left
parts of the phase diagram belong to topological and trivial phases, with
their boundary given by the solid line. The trivial phase enlarges with the
phase boundary shifting rightward as we increase $J$. There are two
topological phases: T-I with $(Q_{1},Q_{2})=(1,0)$ and T-II with $%
(Q_{1},Q_{2})=(0,1)$.

\begin{figure}[t]
\includegraphics[width=1.0\linewidth]{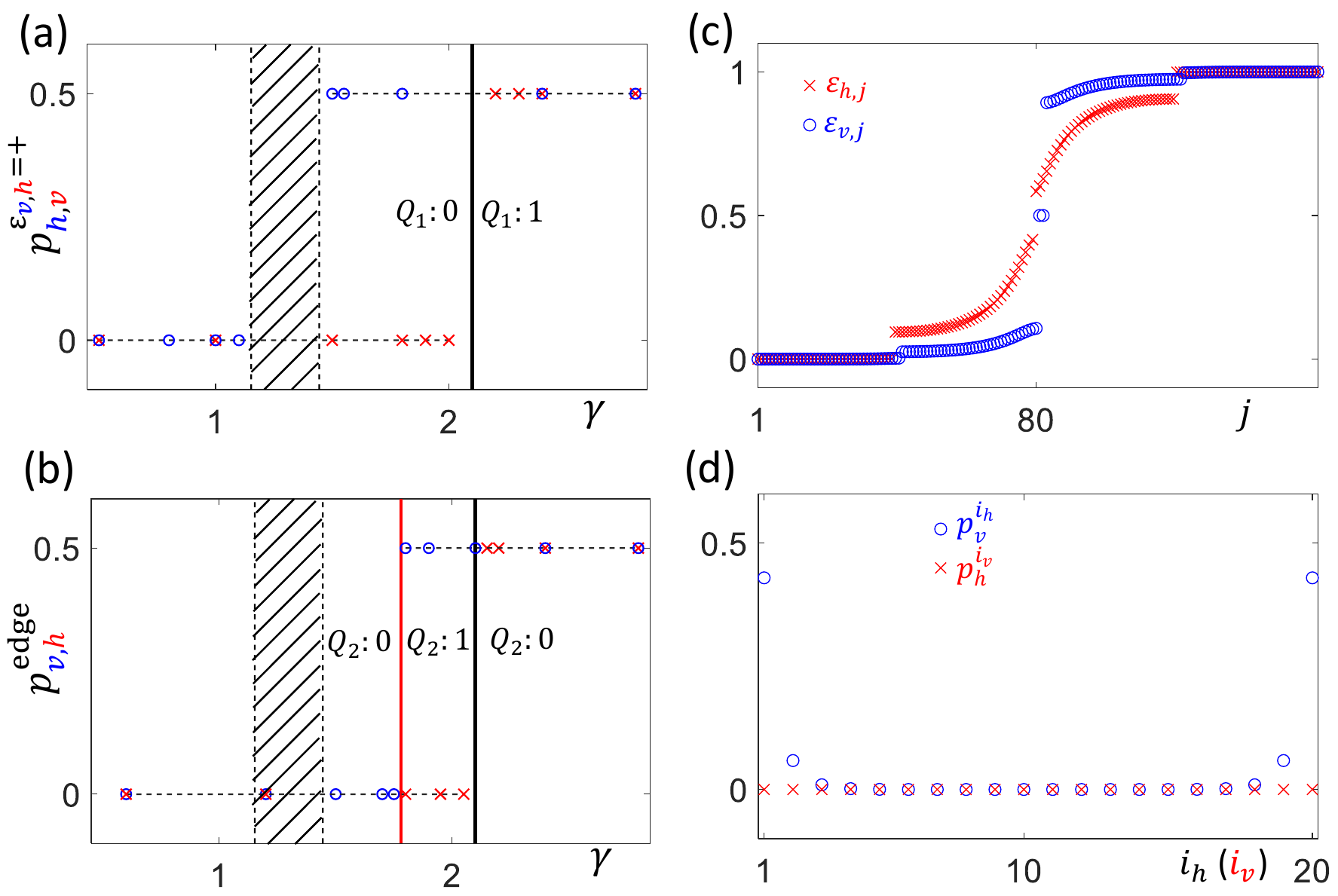}
\caption{(a) and (b) The horizontal (blue circles) and vertical (red
crosses) polarizations for $\protect\lambda _{v}=0.6\protect\lambda _{h}$
[along the thin blue line in Fig.~\protect\ref{fig:phase}(a)]. The
polarization $p_{h}^{\protect\varepsilon _{v}}$ is ill-defined in the
patterned region due to the vanishing gap between Wannier sectors. The phase
boundaries are given by the solid lines. (c) and (d) The Wannier values ($%
\protect\varepsilon _{h,j}$, $\protect\varepsilon _{v,j}$) and
edge-polarization distributions ($p_{h}^{i_{v}}$, $p_{v}^{i_{h}}$) for the
cylindrical geometry with $\protect\lambda _{v}=0.6\protect\lambda _{h}$ and
$\protect\gamma =1.9$ [indicated by the red cross in Fig.~\protect\ref%
{fig:phase}(a)]. Blue circles (red crosses) are the results for open
boundary along horizontal (vertical) direction. The in gap Wannier values at
$\frac{1}{2}$ are responsible for the edge polarization. Other parameters
are the same as in Fig.~\protect\ref{fig:phase}.}
\label{fig:topo_index}
\end{figure}

We first consider the $C_{4}$ symmetric case for $\lambda _{v}=\lambda _{h}$%
, with the open-boundary spectra shown in Fig.~\ref{fig:corner}(a). The
typical Wannier bands for the Hamiltonian Eq.~\ref{Eq:Ham} with periodic
boundaries are shown in Fig.~\ref{fig:phase}(b). There are 8 Wannier bands,
with four located around $\varepsilon =0$, two at $0<\text{Re}[\varepsilon ]<%
\frac{1}{2}$ and two at $0>\text{Re}[\varepsilon ]>-\frac{1}{2}$, forming
three Wannier sectors labeled by $0,\pm $, as shown in Fig.~\ref{fig:phase}%
(b). 
Only the `$\pm $'-Wannier sectors are responsible for the edge topology and
corner states. In fact, the `0'-Wannier sector is trivial in the whole
parameter space and the `$\pm $'-Wannier sectors always have the same
topology. Due to the $C_{4}$ symmetry, we have $p_{h}^{\varepsilon _{v}=\pm
}=p_{v}^{\varepsilon _{h}=\pm }$ and $p_{h}^{\text{edge}}=p_{v}^{\text{edge}%
} $, all of which jump from $0$ to $\frac{1}{2}$ across the phase transition
as $\gamma $ increases [see Fig.~\ref{fig:phase}(c)].
We would like to point out that, with $C_{4}$ symmetry, the topology can
also be characterized by the biorthogonal winding number $W$ along the
high-symmetry line $k_{x}=k_{y}$ in the reflection-rotation ($C_{4}M_{h}$)
subspace~\cite{SM}. 

For $\lambda _{v}\neq \lambda _{h}$, the bulk energy gap persists [see Fig.~%
\ref{fig:phase}(d)], and the phase transitions are driven by gap
close/reopen in the edge spectra and the Wannier bands, which lead to
polarization jumps. In the following, we focus on $\lambda _{v}<\lambda _{h}$
without loss of generality, and show how the topological invariants and
phases change as we increase $\gamma $, as shown in Figs.~\ref%
{fig:topo_index}(a) and (b). 
(i) First, the vertical Wannier bands $\varepsilon _{v,j,\mathbf{k}}$ close
the gap between `0' and `$\pm $' sectors in the patterned region in Fig.~\ref%
{fig:phase}(a)~\cite{SM}. Further increasing $\gamma $ reopens the gap and
leads to the the jump of $p_{h}^{\varepsilon _{v}=\pm }$ from $0$ to $\frac{1%
}{2}$. (ii) Then, the gap for the edge spectra closes and reopens on the red
solid line, where $p_{v}^{\text{edge}}$ jumps from $0$ to $\frac{1}{2}$ and
the system enters the T-II phase with the emergence of corner states. Shown
in Figs.~\ref{fig:topo_index}(c) and (d) are the Wannier values ($%
\varepsilon _{h,j}$ and $\varepsilon _{v,j}$) and edge-polarization
distribution ($p_{h}^{i_{v}}$ and $p_{v}^{i_{h}}$) for the phase T-II on a
cylinder. (iii) Finally, $\varepsilon _{h,j,\mathbf{k}}$ close the gap
between `+' and `$-$' sectors on the black solid line, where both $%
p_{v}^{\varepsilon _{h}=\pm }$ and $p_{h}^{\text{edge}}$ jump from $0$ to $%
\frac{1}{2}$, and we reach the T-I phase. Both T-I and T-II phases support
corner states, and they are distinguished by the edge topology~\cite{SM}.
The T-II phase region shrinks to zero as $\lambda _{v}$ approaches $\lambda
_{h}$, where all edge and Wannier-sector polarizations jump at the same $%
\gamma $ due to the $C_{4}$ symmetry. These phenomena are very different
from the Hermitian case. Especially, one can only have the topological phase
T-I for the Hermitian limit, where all edge polarizations must vanish as
long as $Q_{1}=0$~\cite{PhysRevB.96.245115}. The appearance of phase T-II is
a result of the interplay between the non-Hermiticity and the $C_{4}$
symmetry breaking~\cite{SM}.

\emph{\textcolor{blue}{Discussions}.---} It is possible to generalize our
study by considering different flux configurations.
As a simple example, one may consider $\phi =0$ and set $\sigma _{h,v}^{\phi
}=\sigma _{h,v}^{0}$ in the Hamiltonian Eq.~\ref{Eq:Ham}. For such a zero
flux model, the Hermitian part is a gapless metal when $|\lambda
_{h}-\lambda _{v}|\leq 2J$. The gain/loss term effectively reduces the
tunneling $J$ and can open a topological gap with in-gap corner states~\cite%
{SM}. Moreover, 
it is straightforward to generalize our non-Hermitian model and bulk-corner
correspondence to higher-dimensional systems (e.g., $3$D system supporting
third-order topological phases with quantized octupole moment)~\cite{SM}.
Finally, 
we consider a general asymmetric-tunneling model (without on-site gain/loss)
as an example of chiral symmetric systems, and confirm the bulk-corner
correspondence numerically~\cite{SM}. The asymmetric tunnelings break both
the Hermiticity and reflection symmetries (other symmetries like $C_{4}$
rotation or reflection-rotation $C_{4}M_{h}$ are also broken). As we
increase the strength of the non-Hermiticity (i.e., asymmetry), the system
can transform from the trivial phase to the second-order topological phase
with zero-energy modes at four corners, which are characterized by the
non-trivial topology of the biorthogonal nested Wilson loops~\cite{SM}.

\emph{\textcolor{blue}{Conclusion}.---} In summary, we propose a scheme to
realize non-Hermitian higher-order topological insulators by simply
controlling the on-site gain or loss, and show that the non-Hermitian corner
states are characterized by the bulk topology in the form of biorthogonal
nested Wilson loops or edge polarizations. The generalized bulk-corner
correspondence may work for a wide class of non-Hermitian $d$-dimensional $d$%
-order topological systems with reflection or chiral symmetries. The
proposed model can be realized easily in experiments. Our work offers a
tunable method for manipulating corner states through dissipation control,
and paves the way for the study of various non-Hermiticity induced
higher-order topological states of matter and the classifications of them.

\begin{acknowledgments}
\textbf{Acknowledgements}: This work is supported by AFOSR
(FA9550-16-1-0387), NSF (PHY-1806227), and ARO (W911NF-17-1-0128).
Part of C.Z. work was performed at the Aspen Center for Physics, which is supported by National Science Foundation grant PHY-1607611.
\end{acknowledgments}

\newpage
\newpage
\begin{widetext}
\section*{Supplementary Materials}

\setcounter{figure}{0} \renewcommand{\thefigure}{S\arabic{figure}} %
\setcounter{equation}{0} \renewcommand{\theequation}{S\arabic{equation}}


\subsection{Energy band structures}

The Hamiltonian Eq.~1 in the main text has 16 energy bands,
and the typical band structure is shown in Fig.~\ref{FigS:band}. The bands
are two-fold degenerate in both real and imaginary parts. There is an
exceptional loop (or exceptional ring) for the occupied or unoccupied bands,
which is trivial in the sense that a loop surrounding it has zero vorticity $%
\oint \partial _{\mathbf{k}}\text{Arg}[E(\mathbf{k})]d\mathbf{k}=0$~\cite%
{PhysRevLett.120.146402S}. In the calculation of the Wannier bands, the
exceptional loop is excluded which does not affect the integral because they
are only two points in the horizontal or vertical direction. The numerical
band structures show that the band gap is minimized at $\mathbf{k}=0$, which
should be the band touching point if there are any gap closing. The energy
bands have analytic expressions at $\mathbf{k}=0$, which are given by (all
bands are two-fold degenerate)
\begin{equation}
E=\pm \sqrt{2J^{2}+\lambda _{h}^{2}+\lambda _{v}^{2}-\gamma ^{2}\pm 2\sqrt{%
(J^{2}-\gamma ^{2})(\lambda _{h}^{2}+\lambda _{v}^{2})\pm 2J^{2}\lambda
_{h}\lambda _{v}}}.  \label{Eq:Energy}
\end{equation}%
For the gap closing at zero energy $E=0$, we have the only solution $\lambda
_{h}=\lambda _{v}$ and $\gamma =\sqrt{2(J^{2}-\lambda _{h}^{2})}$. As a
result, the bulk band gap persists across the phase transition for $\lambda
_{h}\neq \lambda _{v}$, while the gap closes at the phase transition point $%
\gamma =\gamma _{c}\equiv \sqrt{2(J^{2}-\lambda _{h}^{2})}$ for $\lambda
_{h}=\lambda _{v}$.

Although our system has no parity-time (PT) symmetry for $\lambda
_{h},\lambda _{v}\neq 0$, we find that when $\gamma $ is small, half of the
eigenenergies are real inside the exceptional ring. As we increase $\gamma $%
, the exceptional ring shrinks to the point $\mathbf{k}=(0,0)$ and
disappears when $\gamma >{J(\lambda _{h}+\lambda _{v})}/\sqrt{\lambda
_{h}^{2}+\lambda _{v}^{2}}$, where Im$[E(\mathbf{k})]$ opens a gap. For the
parameters in Figs.~2(a)(b) of the main text, Im$[E(\mathbf{k})]$ opens a
gap at $\gamma \simeq 1.94$. Since the exceptional ring corresponds to
degeneracy within the occupied (or unoccupied) bands Re$[E(\mathbf{k})]<0$
(or Re$[E(\mathbf{k})]>0$), it does not affect the topology for the gap at Re%
$[E(\mathbf{k})]=0$. The exceptional ring can be understood by noticing the
pseudo-Hermiticity of the Hamiltonian in Eq.~1 of the main text, that is, $%
\chi _{h}H(\mathbf{k})\chi _{h}^{-1}=H^{\dag }(\mathbf{k})$ and $\chi _{v}H(%
\mathbf{k})\chi _{v}^{-1}=H^{\dag }(\mathbf{k})$ with $\chi _{h}=\exp (ik_{x}%
{\tau _{h}^{z}}/{2})\tau _{h}^{x}$ and $\chi _{v}=\exp (ik_{x}{\tau _{v}^{z}}%
/{2})\tau _{v}^{x}$. The pseudo-Hermiticity guarantees that for $%
H|u_{m}\rangle =E_{m}|u_{m}\rangle $, one has $E_{m}^{\ast }\langle
u_{m}|\chi _{h}|u_{m}\rangle =E_{m}\langle u_{m}|\chi _{h}|u_{m}\rangle $
and $E_{m}^{\ast }\langle u_{m}|\chi _{v}|u_{m}\rangle =E_{m}\langle
u_{m}|\chi _{v}|u_{m}\rangle $, which means that the spectrum $%
E_{m}=E_{m}^{\ast }$ is real when either $\langle u_{m}|\chi
_{h}|u_{m}\rangle \neq 0$ or $\langle u_{m}|\chi _{v}|u_{m}\rangle \neq 0$.
Outside the exceptional ring, the pseudo-Hermiticity is fully broken, i.e., $%
\langle u_{m}|\chi _{h}|u_{m}\rangle =\langle u_{m}|\chi _{h}|u_{m}\rangle
=0 $ for all states $m$. While inside the exceptional ring, the
pseudo-Hermiticity is partially broken, namely, $\langle u_{m}|\chi
_{h}|u_{m}\rangle =\langle u_{m}|\chi _{h}|u_{m}\rangle =0$ for one half of
the states $m$, and the other half satisfy either $\langle u_{m}|\chi
_{h}|u_{m}\rangle \neq 0$ or $\langle u_{m}|\chi _{v}|u_{m}\rangle \neq 0$.
For strong enough gain/loss rate, the exceptional ring disappears with gap
opening in Im$[E(\mathbf{k})]$, where the pseudo-Hermiticity is fully broken
in the whole Brillouin zone.

\begin{figure}[htb]
\includegraphics[width=0.6\linewidth]{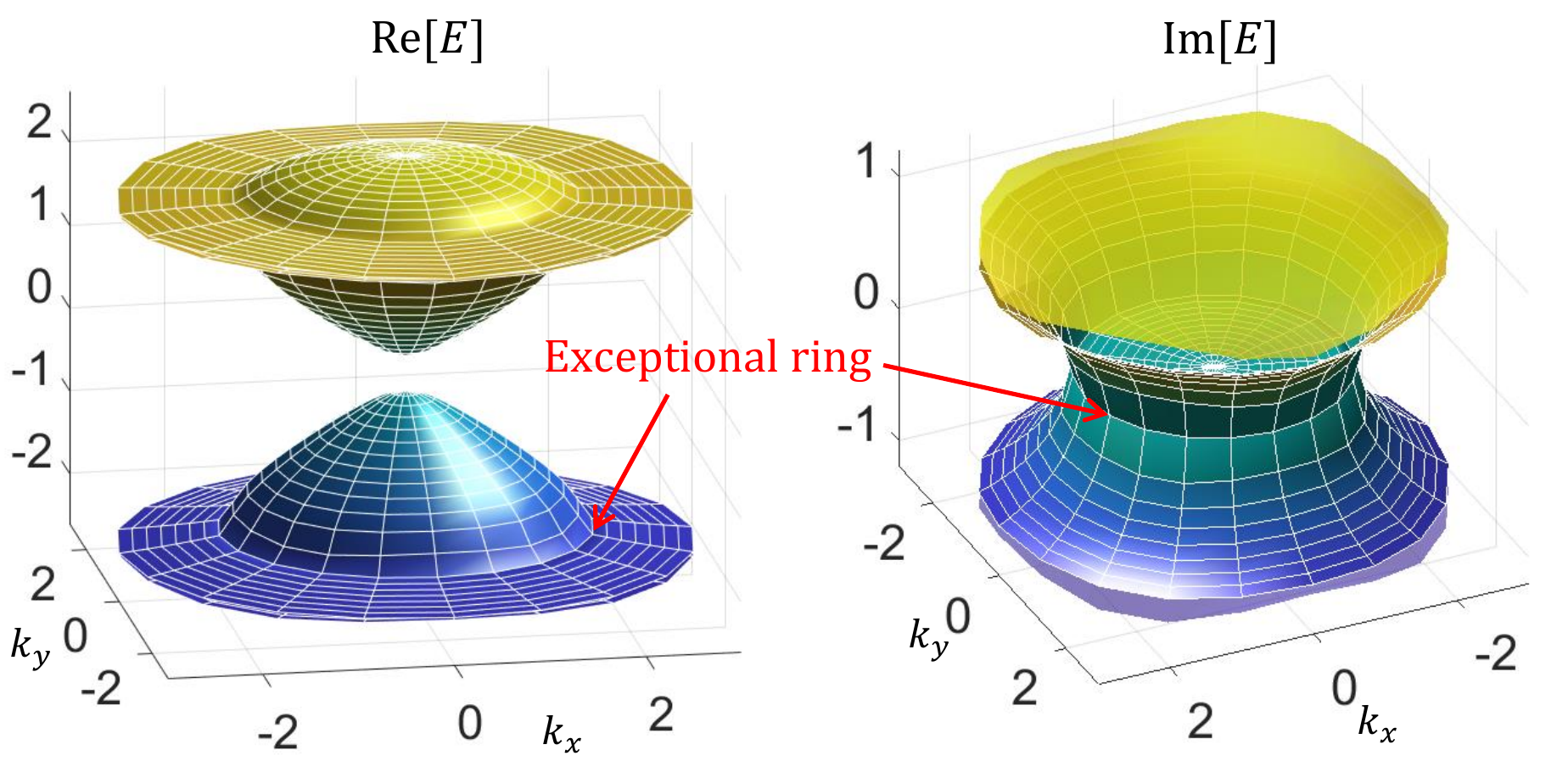}
\caption{Typical band structure (Left panel: real parts; right panel:
imaginary parts) of the Hamiltonian in the main text with periodic
boundaries. Notice that the top transparent yellow band in the imaginary
part corresponds to that breaking pseudo-Hermiticity even inside the
exceptional ring. Shown in the plots are the $C_{4}$ symmetric case with $%
\protect\lambda _{v}=\protect\lambda _{h}$, $J=\protect\sqrt{2}\protect%
\lambda _{h}$ and $\protect\gamma =1.6\protect\lambda _{h}$. We set $\protect%
\lambda _{h}=1$ as the energy unit. }
\label{FigS:band}
\end{figure}



\subsection{Topological invariants and phases}

\emph{Biorthogonal nested Wilson loop.} In the main text, we define the
biorthogonal nested Wilson loop, based on which we obtain the topological
invariants. In particular, the biorthogonal nested Wilson loop is defined on
the non-Hermitian Wannier band basis
\begin{equation}
|w_{h,j,\mathbf{k}}^{R(L)}\rangle =\sum_{m=1}^{N_{\text{occ}}}|u_{m,\mathbf{k%
}}^{R(L)}\rangle \lbrack |\varepsilon _{h,j,\mathbf{k}}^{R(L)}\rangle ]_{m}%
\text{,}  \label{Eq:Wannier_basis}
\end{equation}%
with $\langle w_{h,j,\mathbf{k}}^{L}|w_{h,j^{\prime },\mathbf{k}}^{R}\rangle
=\delta _{j,j^{\prime }}$ and $N_{\text{occ}}$ the number of occupied bands.
The biorthogonal nested Berry connection in the vertical direction can be
defined as $\tilde{A}_{v}^{\varepsilon _{h}}(\mathbf{k})=-i\langle w_{j,h,%
\mathbf{k}}^{L}|\partial _{k_{y}}|w_{j^{\prime },h,\mathbf{k}}^{R}\rangle $
with $j$ and $j^{\prime }$ running over the Wannier band sector $\varepsilon
_{h}$. The corresponding biorthogonal nested Wilson loop is%
\begin{equation}
\tilde{W}_{h,\mathbf{k}}=\mathcal{P}\exp [i\int_{k_{y}}^{k_{y}+2\pi }\tilde{A%
}_{v}^{\varepsilon _{h}}(k_{x},k_{y}^{\prime })dk_{y}^{\prime }]\text{,}
\label{Eq:Nested Wilson}
\end{equation}

\emph{Edge polarizations.} For Hermitian systems with trivial Wannier bands,
it was shown that the topological invariant is characterized by the edge
polarizations~\cite{PhysRevB.96.245115S}. We find that this can also be
applied to our non-Hermitian system by calculating edge polarizations in the
biorthogonal basis. In particular, we consider a cylindrical geometry with
open boundary along vertical direction, and treat the system as a
pseudo-one-dimensional system along the horizontal direction with $N_{\text{%
occ}}\times N_{v}$ occupied bands ($N_{v}$ is number of unit cells along the
vertical direction). Similar to the torus case with a fixed $k_{y}$ (as
described in the main text), we can obtain the biorthogonal Wilson loop $%
\left[ W_{h,k_{x}}\right] _{m,n}$, and the Wannier bands $\varepsilon _{h,j}$
with $m,n,j\in 1,2,...,N_{\text{occ}}\times N_{v}$. We define the horizontal
polarization as a function of vertical site index $i_v$ as
\begin{equation}
p_{h}^{i_{v}}=\frac{1}{2\pi }\int dk_{x}\sum_{j,n,m,l}\varepsilon
_{h,j}\cdot \left[ \langle u_{n,k_{x}}^{L}|\right] _{i_{v},l}[\langle
\varepsilon _{h,j}^{L}|]_{n}\cdot \left[ |u_{m,k_{x}}^{R}\rangle \right]
_{i_v,l}[|\varepsilon _{h,j}^{R}\rangle ]_{m},  \label{eqS:edgeP}
\end{equation}%
where $l$ is the orbital index in the unit cell $i_{v}$, $m,n$ are the
occupied band indexes, and $j$ is the Wannier band index.
Similarly, we may consider open boundary along the horizontal direction and
obtain the vertical polarization $p_{v}^{i_{h}}$ as a function of horizontal
unit-cell index $i_{h}$. The edge polarization $p_{h(v)}^{\text{edge}}$ is
given by the summation of $p_{h}^{i_{v}}$ ($p_{v}^{i_{h}}$) near one edge~%
\cite{PhysRevB.96.245115S}. 

\emph{Winding number.} In the presence of additional $C_{4}$ symmetry,
the Hermitian part of the Hamiltonian in the main text can be characterized
by the winding number of its projection $H_{\pm }(\mathbf{k})$ onto the
reflection-rotation ($C_{4}M_{h}$) subspace along the high-symmetry line $%
k_{x}=k_{y}=k$ [$H_{\pm }(k,k)$ acts on the subspace satisfying $%
C_{4}M_{h}=\pm 1$]~\cite{Imhof2018Topolectrical-circuitS}. For the
non-Hermitian system, the winding number is evaluated on the biorthogonal
basis, leading to the following biorthogonal winding number
\begin{equation}
W=-\frac{i}{\pi }\text{Tr}\left[ \mathcal{P}\oint \langle u_{\pm
,m}^{L}|\partial _{k_{y}}|u_{\pm ,n}^{R}\rangle dk\right] ,
\label{Eq:Winding}
\end{equation}%
where $(m,n)$ runs over the occupied bands of $H_{\pm }(k,k)$ and $|u_{\pm
,n}^{R,L}\rangle $ are the corresponding right and left eigenstates.

\emph{Topological phases.} 
For $\lambda _{v}\neq \lambda _{h}$, we have $p_{h}^{\varepsilon _{v}=\pm
}\neq p_{v}^{\varepsilon _{h}=\pm }$ and $p_{h}^{\text{edge}}\neq p_{v}^{%
\text{edge}}$, and these polarizations jump at different loss rate $\gamma $
due to the lack of $C_{4}$ symmetry. We focus on $\lambda _{v}<\lambda _{h}$
in the following (for $\lambda _{v}>\lambda _{h}$, the physics is similar
except that the horizontal and vertical directions exchange their roles). In
the patterned region in Fig.~3(a) of the main text, the vertical Wannier
bands $\varepsilon _{v,j,\mathbf{k}}$ close the gap between `0' and `$\pm $'
sectors, as shown in Fig.~\ref{FigS:Wannier}(a). The horizontal Wannier
bands $\varepsilon _{h,j,\mathbf{k}}$ close and reopen the gap between `+'
and `$-$' sectors at the phase boundary between topological phases T-I ($%
Q_{1}=1$, $Q_{2}=0$) and T-II ($Q_{1}=0$, $Q_{2}=1$), as shown in Fig.~\ref%
{FigS:Wannier}(b). For other $\gamma $, the Wannier bands are gapped. As an
example, we plot the typical Wannier bands for the T-I phase in Figs.~\ref%
{FigS:Wannier}(c) and (d). The phase boundary between the trivial phase ($%
Q_{1}=0$, $Q_{2}=0$) and the topological phase T-II is characterized by the
polarization jump only for $Q_{2}$ (i.e., $p_{v}^{\text{edge}}$), which is
induced by the gap close/reopen in the edge spectra. Both $Q_{1}$ and $Q_{2}$
jump at the phase boundary between T-II and T-I, and both $p_{h,v}^{\text{%
edge}}$ and $p_{h,v}^{\varepsilon _{v,h}=\pm }$ become non-trivial (i.e.,
equals to $\frac{1}{2}$) in the phase T-I. Near the phase boundary, $%
p_{h}^{i_{v}}$ exponentially penetrates into the bulk, therefore we need to
consider a large system to obtain the quantized edge polarization. The
global polarization [\textit{i.e.}, the summation of $p_{h}^{i_{v}}$ ($%
p_{v}^{i_{h}}$) over all unit cell $i_{v}$ ($i_{h}$)] is always zero (mod 1).

These features are quite different from the Hermitian case, where all edge
polarizations must be zero as long as $Q_{1}=0$. The topological phase T-II
is a result of the interplay between the non-Hermiticity and the $C_{4}$
symmetry breaking. Even though the Wannier-sector polarization is trivial
along vertical direction in phase T-II, the non-Hermitian particle loss can
induce non-trivial vertical polarization on the horizontal edge (\textit{i.e.%
}, $i_{h}=1$ and $i_{h}=N_{h}$). In particular,
the gapped edge states first appear on the horizontal boundaries due to the
stronger horizontal coupling (i.e., $\lambda _{h}>\lambda _{v}$), then the
particle loss $\gamma $ drives the phase transition of these edge states
(from trivial phase $p_{v}^{\text{edge}}=0$ to T-II phase $p_{v}^{\text{edge}%
}=\frac{1}{2}$) prior to the jump of $Q_{1}$. Upon the transition from
trivial phase to T-II phase, the edge-state gaps close and reopen.

As $\gamma $ increases further, the system enters the T-I phase with the
appearance of fully separated edge states on both the horizontal and
vertical boundaries (There are no fully separated edge states on the
vertical edges in the trivial and T-II phases). In particular, we consider
periodic (open) boundary condition in horizontal (vertical) direction. The
edge spectrum exists for all momentum $k_{x}$ and is fully separated from
the bulk spectrum in phase T-I; while in phase T-II, the edge spectrum
merges into the bulk at certain momentum $k_{x}$ and disappears for a finite
momentum interval (see Fig.~\ref{FigS:edgespectrum}). The change of edge
spectrum leads to the change of edge polarization at the phase boundary
between T-I and T-II phases.

\emph{Symmetry breaking perturbations.} In general, higher-order topological
phases are protected by symmetries~\cite{PhysRevB.96.245115S}. As a result,
when both reflection and chiral symmetries are broken, both the multipole
moments and the biorthogonal topological invariants are no longer quantized.
In particular, if we introduce perturbations (e.g., $\delta \sigma
_{h}^{z}\sigma _{v}^{z}$) such that the system breaks both reflection and
chiral symmetries, the four corner states break their degeneracy and shift
toward the bulk as the perturbation strength increases~\cite%
{PhysRevB.96.245115S}. Moreover, the biorthogonal nested-Wilson-loop and
edge-polarization theory does not require additional symmetries such as the $%
C_{4}$ rotational or reflection-rotation symmetries.

\begin{figure}[tb]
\includegraphics[width=0.6\linewidth]{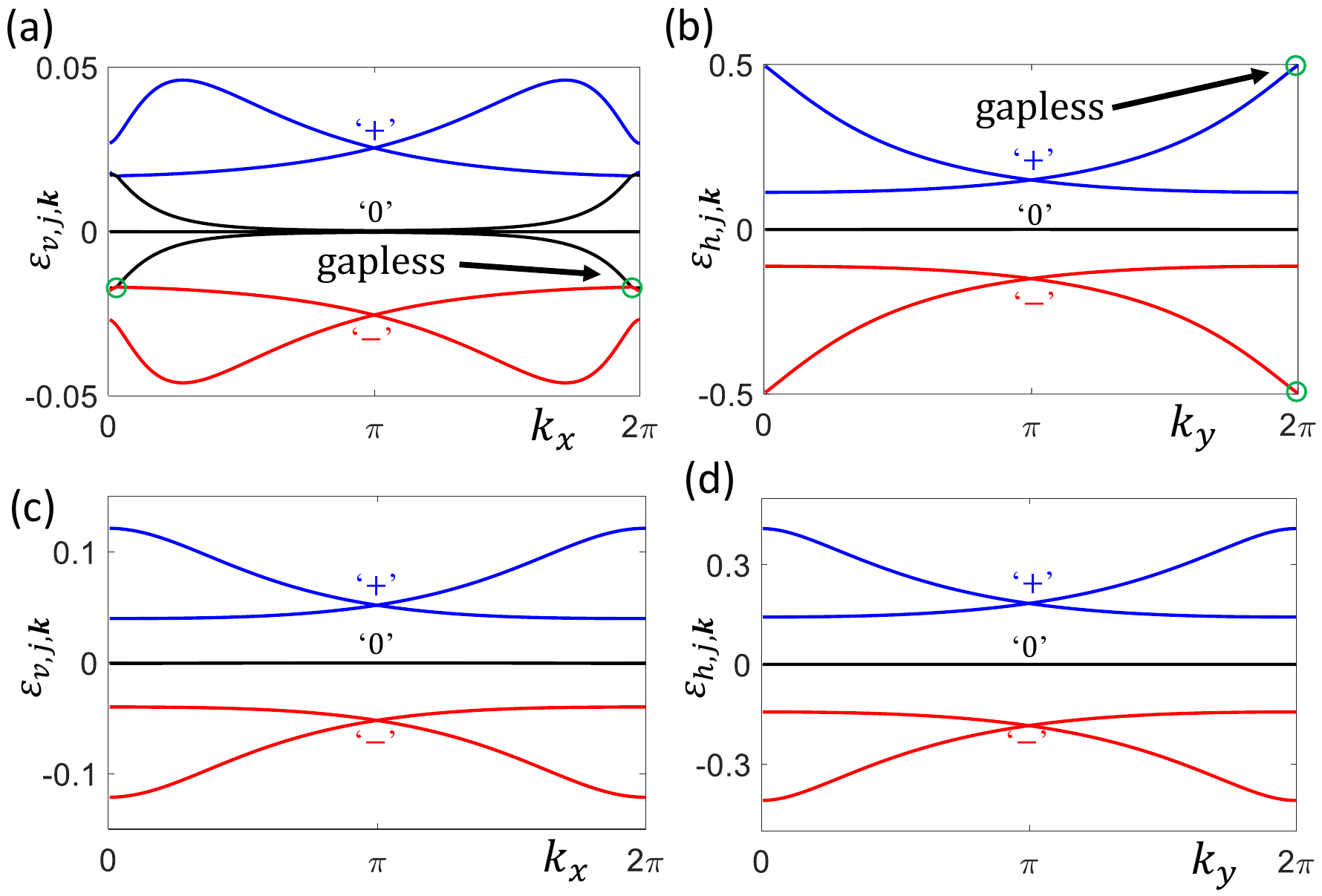}
\caption{(a) Vertical Wannier band gap closing in the patterned region in
Fig.~3(a) in the main text, with $\protect\gamma =\protect\sqrt{2}$. The
horizontal Wannier band is gapped. (b) Horizontal Wannier band gap closing
on the boundary between T-I and T-II with $\protect\gamma =2.1$. The
vertical Wannier band is gapped. (c) and (d) Horizontal and vertical Wannier
band structures with $\protect\gamma =2.5\protect\lambda _{h}$. The
imaginary parts of the Wannier bands are locked at zero. Common parameters: $%
\protect\lambda _{v}=0.6\protect\lambda _{h}$, $\protect\lambda _{h}=1$ is
the energy unit.}
\label{FigS:Wannier}
\end{figure}

\begin{figure}[tb]
\includegraphics[width=0.8\linewidth]{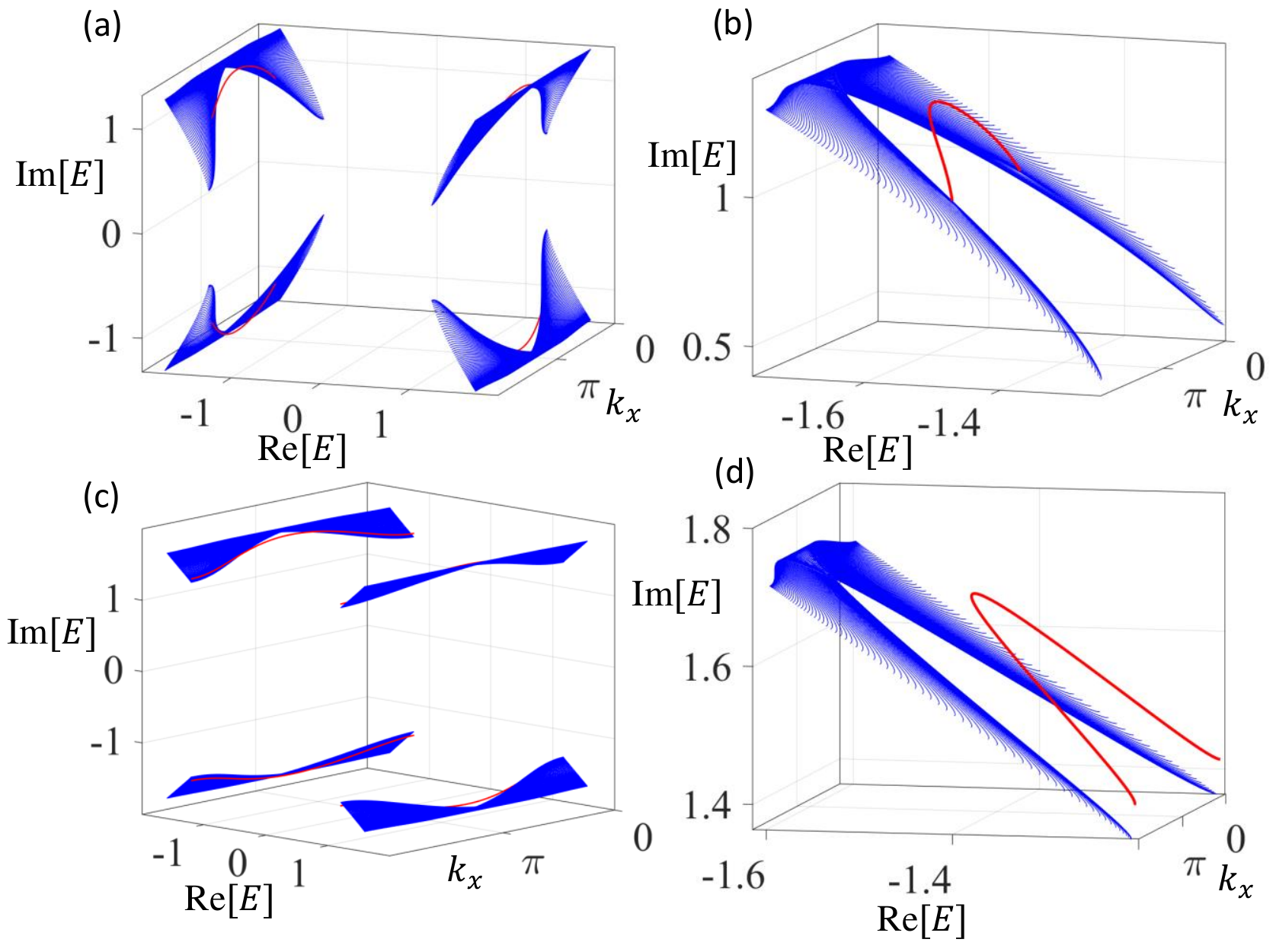}
\caption{(a) Spectrum of the T-II phase with periodic (open) boundary
condition in horizontal (vertical) direction and $\protect\gamma =2$. The
red lines show the edge spectrum on the open vertical boundary which merge
into the bulk at certain $k_{x}$. (b) The zoom in of (a). (c) Spectrum of
the T-I phase with periodic (open) boundary condition in horizontal
(vertical) direction and $\protect\gamma =2.4$. The red lines show the edge
spectrum on the open vertical boundary. (d) The zoom in of (c). The spectrum
is symmetric with respect to Re$[E]=0$ and Im$[E]=0$ due to the
pseudo-anti-Hermiticity and pseudo-Hermiticity. Common parameters: $\protect%
\lambda _{v}=0.6\protect\lambda _{h}$, $\protect\lambda _{h}=1$ is the
energy unit.}
\label{FigS:edgespectrum}
\end{figure}

\subsection{Zero flux $\protect\phi=0$ case}

For the zero flux case, the Hamiltonian becomes
\begin{eqnarray}
H(\mathbf{k}) &=&J_{h}\sigma _{h}^{x}+J_{v}\sigma _{v}^{x}\sigma _{h}^{0}+%
\frac{i\gamma }{2}\sigma _{h}^{z}\sigma _{v}^{z}\tau _{h}^{z}\tau _{v}^{z}
\nonumber \\
&&+\lambda _{h}(\tau _{h}^{-}\sigma _{h}^{+}+e^{-ik_{x}}\tau _{h}^{-}\sigma
_{h}^{-}+h.c.)  \label{Eq:Ham0} \\
&&+\lambda _{v}\sigma _{h}^{0}(\tau _{v}^{-}\sigma _{v}^{+}+e^{-ik_{y}}\tau
_{v}^{-}\sigma _{v}^{-}+h.c.).  \nonumber
\end{eqnarray}%
Here we consider $\lambda _{v}>\lambda _{h}\geq J$. When $\gamma =0$ and $%
|\lambda _{h}-\lambda _{v}|\leq 2J$, the Hamiltonian is in the gapless metal
phase~\cite{PhysRevB.96.245115S}. As $\gamma $ increases (the tunneling $J$
is effectively reduced), a topological gap opens with the emergency of
in-gap corner states [see Figs.~\ref{FigS:edgeP} (a) and (b)]. $\lambda
_{h}\neq \lambda _{v}$ is required for the appearance of the gap. Our
numerical results show that the gap opens at momentum $\mathbf{k}=0$, where
we have analytic solutions for the eigenenergies: $E(\mathbf{k}=0)=\pm
\lambda _{v}\pm \lambda _{h}\pm i\gamma $ and $E(\mathbf{k}=0)=\pm \lambda
_{v}\pm \lambda _{h}\pm \sqrt{4J^{2}-\gamma ^{2}}$. The gap opens when $%
\lambda _{v}-\lambda _{h}-\sqrt{4J^{2}-\gamma ^{2}}=0$, leading to the
critical loss rate $\gamma _{c}=\sqrt{4J^{2}-(\lambda _{v}-\lambda _{h})^{2}}
$.

In the metal phase, these is no well defined topological invariant. Even if
there exist zero-energy corner states, they are embedded in the bulk spectra
(no gap protection) and cannot be distinguished. Unfortunately, the
topological invariant of such a model cannot be extracted from the
biorthogonal nested Wilson loop because all the Wannier bands are locked at $%
0$ or $\frac{1}{2}$ with trivial Wannier-sector polarizations. For Hermitian
systems with trivial flat Wannier bands, the second-order topology is
characterized by the biorthogonal edge polarizations~\cite%
{PhysRevB.96.245115S}. In Figs.~\ref{FigS:edgeP} (c)-(f), we show the
Wannier bands $\varepsilon _{h,j}$ and $\varepsilon _{v,j}$, as well as the\
polarization $p_{h}^{i_{v}}$ and $p_{v}^{i_{h}}$ in the topological
insulator phase at a large $\gamma $. We see that the polarization is well
localized at the edges with vanishing bulk distributions, leading to the
non-trivial topological invariant $Q_{2}=1$.

\begin{figure}[t]
\includegraphics[width=0.8\linewidth]{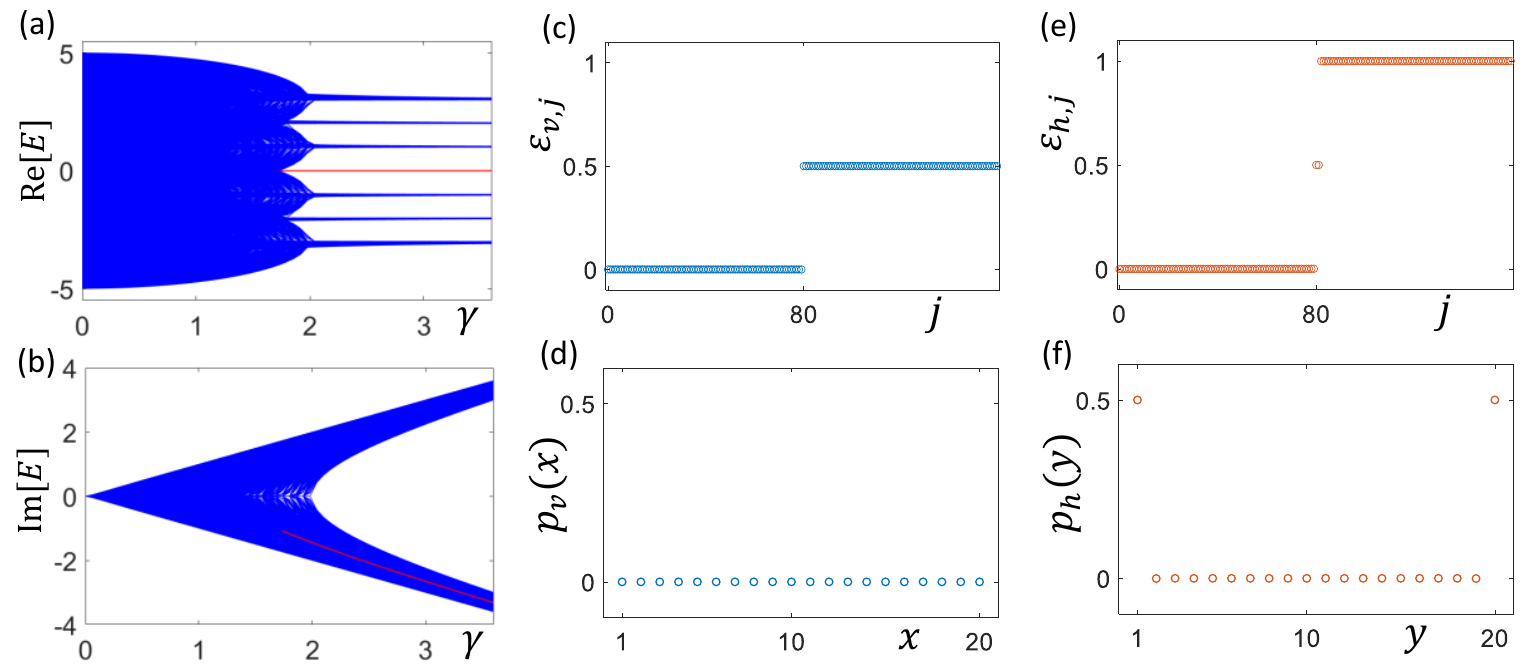}
\caption{(a) and (b) Complex band structures of the Hamiltonian in Eq.~%
\protect\ref{Eq:Ham0} as functions of $\protect\gamma $ with open boundaries
in both directions. (c) and (d) The Wannier bands and polarizations with
periodic (open) boundary in vertical (horizontal) direction. (e) and (f) The
same as in (c) and (d) except that the horizontal (vertical) direction is
periodic (open). $\protect\gamma =2$ in (b)-(f). We use 20 unit cells in the
open direction in all calculations. Common parameters: $J=\protect\lambda %
_{h}$, $\protect\lambda _{v}=2\protect\lambda _{h}$, and $\protect\lambda %
_{h}=1$ is the energy unit. The gap opening point is at $\protect\gamma _{c}=%
\protect\sqrt{4J^{2}-(\protect\lambda _{v}-\protect\lambda _{h})^{2}}=%
\protect\sqrt{3}$.}
\label{FigS:edgeP}
\end{figure}

\begin{figure}[t]
\includegraphics[width=0.8\linewidth]{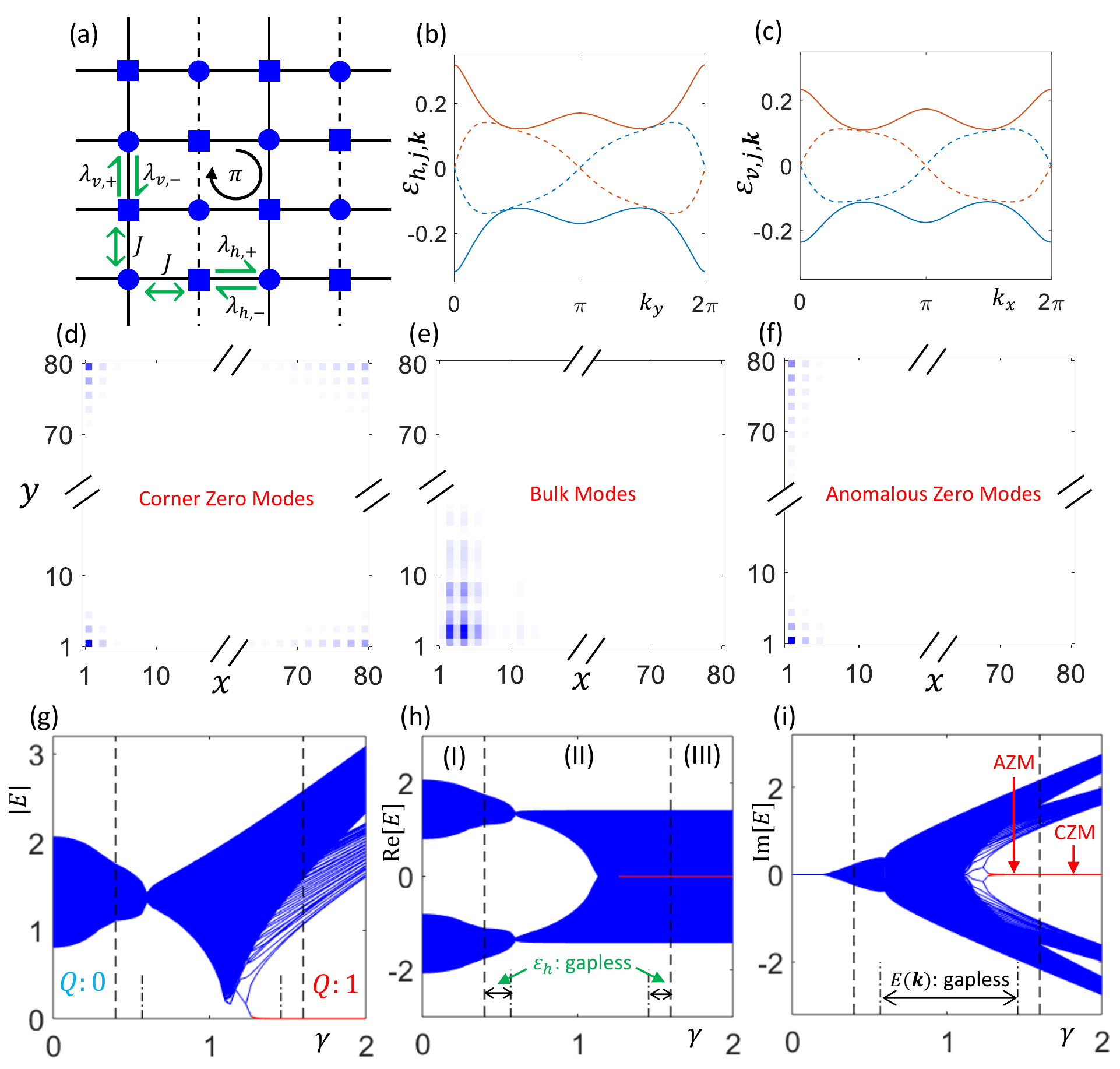}
\caption{(a) Lattice representation of the Hamiltonian in Eq.~\protect\ref%
{Eq:HamC}. (b) and (c) Complex Wannier bands. The red lines are the real
(solid line) and imaginary (dashed line) parts of one Wannier band, while
the blue lines are for the other band. The parameters are: $\protect\lambda %
_{h}=0.3$, $\protect\lambda _{v}=0.6$ and $\protect\gamma =1.8$. (d) and (e)
Density distributions of the corner zero modes (CZMs) and bulk modes for $%
40\times 40$ unit cells with parameters the same as in (b). (f) Density
distributions of the anomalous zero modes (AZMs) in phase (II) with $\protect%
\gamma =1.5$ and other parameters the same as in (b). (g)-(i) Band
structures of the Hamiltonian in Eq.~\protect\ref{Eq:HamC} as functions of $%
\protect\gamma $ with open boundaries in both directions for $40\times 40$
unit cells. Here we set $J=1$ as the energy unit.}
\label{FigS:chiral}
\end{figure}

\subsection{Chiral symmetric model}

As we discussed in the main text, the biorthogonal topological invariant not
only applies to reflection symmetric systems, but also to chiral symmetric
systems with non-Hermiticity induced by asymmetric tunnelings. Because of
the chiral symmetry, the occupied and unoccupied energy bands have the same
Wannier bands (values), and their total summation should be flat and locked
at $0$ mod $1$. Thus, the Wannier bands (values) for occupied energy bands
should be either flat bands locked at $0$ or $\frac{1}{2}$, or appear in $%
\pm \varepsilon $ pairs. Interestingly, we find that the biorthogonal
Wannier-sector or edge polarization
is also quantized to $0$ or $\frac{1}{2}$ mod $1$. This might be understood
by considering the projection (both occupied and unoccupied bands) onto the
Wannier basis as a smooth mapping to an effective one-dimensional model. The
biorthogonal polarization 
should be quantized due to the chiral symmetry and the unoccupied bands give
the chiral partner of the occupied bands.

As an example, we consider a similar asymmetric-tunneling model as that in
Ref.~\cite{arXiv1810.04067S}. The chiral symmetric Hamiltonian under
periodic boundary is
\begin{eqnarray}
H(\mathbf{k}) &=&\sigma _{h}^{+}(J+\lambda _{h,+}e^{ik_{x}})+\sigma
_{h}^{-}(J+\lambda _{h,-}e^{-ik_{x}})  \nonumber \\
&&\sigma _{h}^{z}\sigma _{v}^{+}(J+\lambda _{v,+}e^{ik_{y}})+\sigma
_{h}^{z}\sigma _{v}^{-}(J+\lambda _{v,-}e^{-ik_{y}}),  \label{Eq:HamC}
\end{eqnarray}%
where $\mathbf{\sigma }_{h,v}$ are the Pauli matrices for the degrees of
freedom spanned by circle and square sites, and the inter-cell tunneling is
asymmetric with $\lambda _{v,\pm }=\lambda _{v}\pm \gamma $, $\lambda
_{h,\pm }=\lambda _{h}\pm \gamma $ and $\gamma >0$, as shown in Fig.~\ref%
{FigS:chiral} (a). The chiral symmetry is given by $\Xi =\sigma
_{h}^{z}\sigma _{v}^{z}$, which flips the sign of all square sites, leading
to $\Xi H(\mathbf{k})\Xi ^{-1}=-H(\mathbf{k})$. The (two-fold degenerate)
eigenenergies are
\begin{equation}
E(\mathbf{k})=\pm \sqrt{\lambda _{h}^{2}+\lambda _{v}^{2}+2J^{2}-2\gamma
^{2}+2J\lambda _{h}\cos (k_{x})+2J\lambda _{v}\cos (k_{y})+2iJ\gamma \lbrack
\sin (k_{x})+\sin (k_{y})]},
\end{equation}%
which is gapless in the region $\sqrt{\frac{(J-\lambda _{h})^{2}+(J-\lambda
_{v})^{2}}{2}}<\gamma <\sqrt{\frac{(\lambda _{h}+J)^{2}+(\lambda _{v}+J)^{2}%
}{2}}$ and gapped otherwise.

In the Hermitian limit, the system stays in the topological trivial phase in
the region $J>\lambda _{h,v}$ (we assume $J>\lambda _{v}\geq \lambda _{h}$
without loss of generality). We find that a non-zero $\gamma $ not only
breaks the Hermiticity and the reflection symmetry, but also drives the
system to a second-order topological phase with corner states. The
non-trivial topology is characterized by the biorthogonal nested Wilson
loops (we also calculate the biorthogonal edge polarization which either
leads to a trivial topological invariant or becomes ill defined due to the
non-Hermitian skin effect). Three phases are identified for different $%
\gamma $: phase (I) for $\gamma <J-\lambda _{v}$; phase (II) for $J-\lambda
_{v}<\gamma <J+\lambda _{v}$; phase (III) for $\gamma >J+\lambda _{v}$. In
phases (I) and (III), both $E(\mathbf{k})$ and the Wannier bands $%
\varepsilon _{h(v),j,\mathbf{k}}$ [see Figs.~\ref{FigS:chiral} (b) and (c)]
are gapped. We have $Q_{1}=1$ in phase (III) and $Q_{1}=0$ in phase (I).
Therefore phase (III) is topological non-trivial with four zero-energy
corner modes under open boundaries, and they are located at the four
corners, respectively, leading to quantized quadruple moment [see Figs.~\ref%
{FigS:chiral} (d)]. In phase (II), either $E(\mathbf{k})$ or $\varepsilon
_{h(v),j,\mathbf{k}}$ is gapless, and we do not have well defined $Q_{1}$.
In all phases, the bulk states are located at one corner [see Fig.~\ref%
{FigS:chiral} (e)] when open boundaries are considered.

In Figs.~\ref{FigS:chiral} (g)-(i), we plot the complex energies as
functions of $\gamma $ with open boundaries. We notice that the
open-boundary bulk spectra is gapped everywhere (the imaginary energy gap
opens before the real energy gap closes), which are quite different with the
periodic case. Under open boundaries, there might still exist anomalous zero
modes in phase (II); however, the four in-gap states are located at one (or
two) corner(s) [see Fig.~\ref{FigS:chiral} (f)]. Thus they do not correspond
to quantized quadruple moment, which is why we do not have a well-defined $%
Q_{1}$ in phase (II). Nevertheless, the in-gap states in phase (II) might be
characterized by, for example, the non-Bloch theory~\cite%
{PhysRevLett.121.136802S,PhysRevLett.121.086803S,arXiv1810.04067S}, and the
anomalous in-gap states are believed to be a result of the interplay between
skin effect and finite size effect. Developing a general bulk-corner
correspondence for such anomalous zero modes is also very interesting and
can be addressed in future work.


We would like to emphasize that the direct diagonalization of the
Hamiltonian Eq.~\ref{Eq:HamC} with open boundaries may not give the correct
corner states in phase (III). This is because there are couplings
(exponentially weak) between four corner states for a finite system, which
mix the states at four corners. For such a mixed state, the skin effects
wash out the components in three corners. However, we can isolate the state
at each corner by an infinitesimal on-site detuning $\delta \sigma
_{h}^{z}\sigma _{v}^{z}$~\cite{PhysRevB.96.245115S} or by considering open
boundaries with broken unit cell~\cite{PhysRevLett.121.026808S}. The skin
effects only affect the spatial profiles (decay rates) of the corner states
without changing their localizing corner positions in phase (III) with $%
Q_{1}=1$. While in phase (II), the skin effects become strong enough and all
corner states are shifted to a single (or two) corner(s).

\begin{figure}[t]
\includegraphics[width=0.6\linewidth]{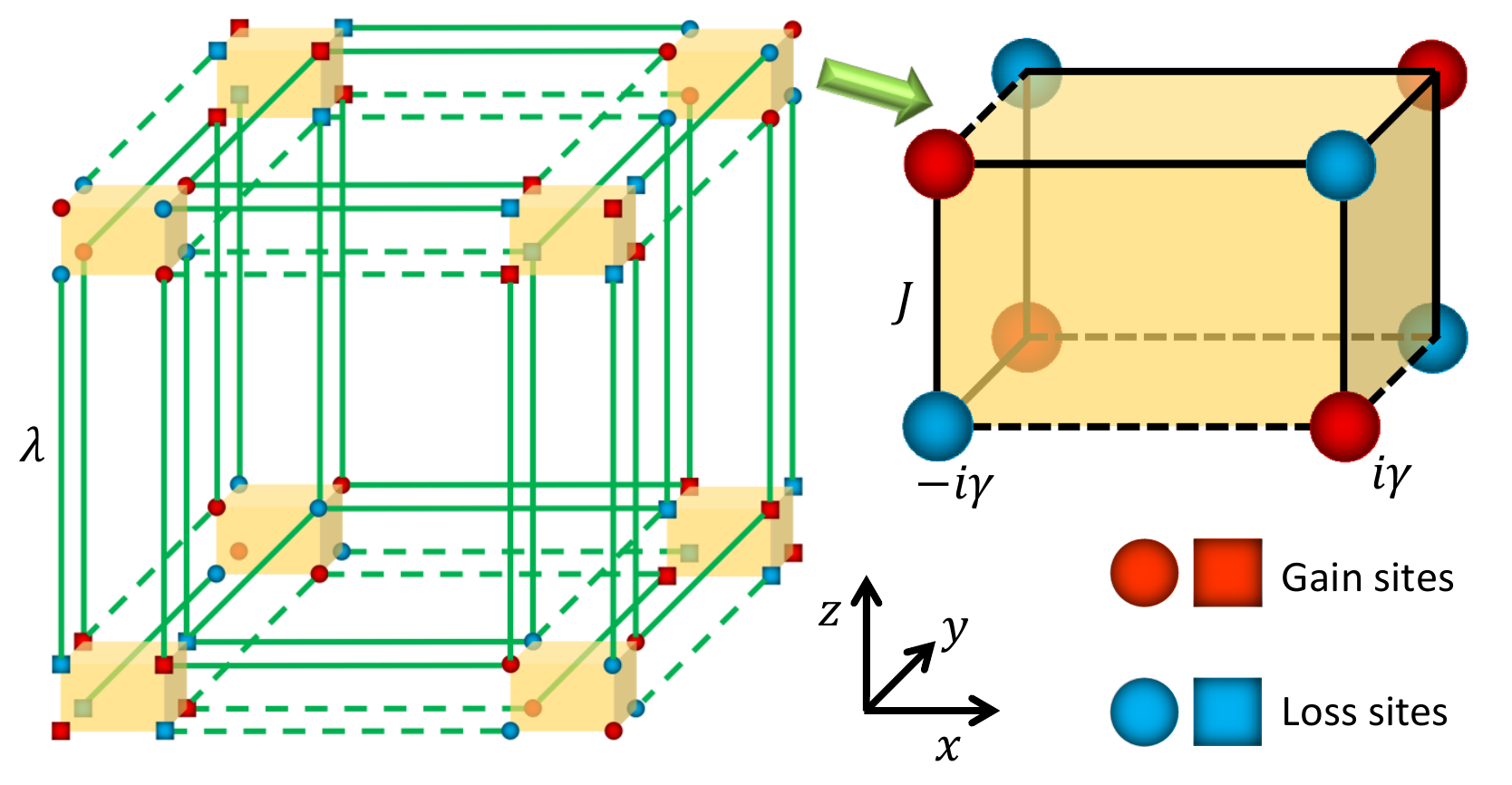}
\caption{Left panel: Lattice representation of the 3D non-Hermitian model
supporting third-order topological phases. Right panel: zoom in of the
building block in the yellow cube (similar for the cube formed by square
sites).}
\label{FigS:3D}
\end{figure}

\subsection{Third-order topological phases in 3D systems}

In this section, we show how to generalize our non-Hermitian model and
bulk-corner correspondence to higher-dimensional systems. As an example, we
consider a $3$D system supporting third-order topological phases with
quantized octupole moment, as shown in Fig.~\ref{FigS:3D}. There is an
effective magnetic flux $\phi =\pi $
for each plaquette, which appears as the tunneling phases on the dashed
lines. The non-Hermiticity is introduced by the particle loss (gain) on all
blue (red) lattice sites with rate $\gamma $. For simplicity, we choose the
tunneling strengths to be the same for all green (black) links and denote
them as $\lambda $ ($J$). The Hamiltonian in momentum space reads
\begin{eqnarray}
H(\mathbf{k}) &=&J\sigma _{a}^{x}+J\sigma _{h}^{x}\sigma _{a}^{z}+J\sigma
_{v}^{x}\sigma _{h}^{z}\sigma _{a}^{z}+i\gamma \sigma _{a}^{z}\sigma
_{h}^{z}\sigma _{v}^{z}\tau _{a}^{z}\tau _{h}^{z}\tau _{v}^{z}+\lambda (\tau
_{a}^{-}\sigma _{a}^{+}+e^{-ik_{z}}\tau _{a}^{-}\sigma _{a}^{-}+h.c.)
\nonumber \\
&&+\lambda \sigma _{a}^{z}(\tau _{h}^{-}\sigma _{h}^{+}+e^{-ik_{x}}\tau
_{h}^{-}\sigma _{h}^{-}+h.c.)+\lambda \sigma _{h}^{z}\sigma _{a}^{z}(\tau
_{v}^{-}\sigma _{v}^{+}+e^{-ik_{y}}\tau _{v}^{-}\sigma _{v}^{-}+h.c.).
\label{Eq:Ham3D}
\end{eqnarray}%
$\mathbf{\sigma }_{a,h,v}$ ($\mathbf{\tau }_{a,h,v}$) are the Pauli matrices
for the degrees of freedom spanned by red and blue (circle and square)
sites, and $a,h,v$ represent the $z$, $x$ and $y$ directions, respectively.

For $J>\lambda $, the system is in a trivial phase, and as we increase the
gain/loss rate $\gamma $, the system undergoes a phase transition to the
third-order topological phase at $\gamma =\gamma _{c}\equiv \sqrt{%
3(J^{2}-\lambda ^{2})}$. In general, we have $\gamma _{c}\equiv \sqrt{%
d(J^{2}-\lambda ^{2})}$ for such $C_{4}$-rotational symmetric case, with $d$
the system dimension. To obtain the biorthogonal topological invariants, we
can define the biorthogonal Wilson loop operator along $k_{x}$, just like
the 2D case. From which we can obtain the 2D Wannier Hamiltonian $%
H_{W_{h}}(k_{y},k_{z})$ and Wannier bands $\varepsilon _{h}(k_{y},k_{z})$ in
the $k_{y}$-$k_{z}$ plane. Then we calculate the 2D biorthogonal topological
invariants (i.e., the quadrupole moment) of the non-Hermitian Hamiltonian $%
H_{W_{h}}(k_{y},k_{z})$ by taking the Wannier sector $\varepsilon
_{h}(k_{y},k_{z})\in (0,\frac{1}{2})$ as the effective \textquotedblleft
occupied bands"~\cite{PhysRevB.96.245115S}. Similarly, the biorthogonal edge
polarizations can be generalized to the biorthogonal corner polarizations.
We may consider periodic boundary along the $x$ direction and open boundary
along the other two directions, then we treat the system as a
pseudo-one-dimensional system and calculate the biorthogonal Wannier values $%
\varepsilon _{h,j}$ along the periodic direction. Using Eq.~\ref{eqS:edgeP},
we can get the $x$-direction polarization as a function of $y$- and $z$%
-direction site index $i_{v},i_{a}$, which should be well localized at the
corners in the $i_{v}$-$i_{a}$ plane for the third-order topological phases.
Moreover, one may also calculate the edge polarization of the 2D Wannier
Hamiltonian $H_{W_{h}}(k_{y},k_{z})$, which should be non-trivial for the
third-order topological phases~\cite{PhysRevB.96.245115S}.

\begin{figure}[t]
\includegraphics[width=0.6\linewidth]{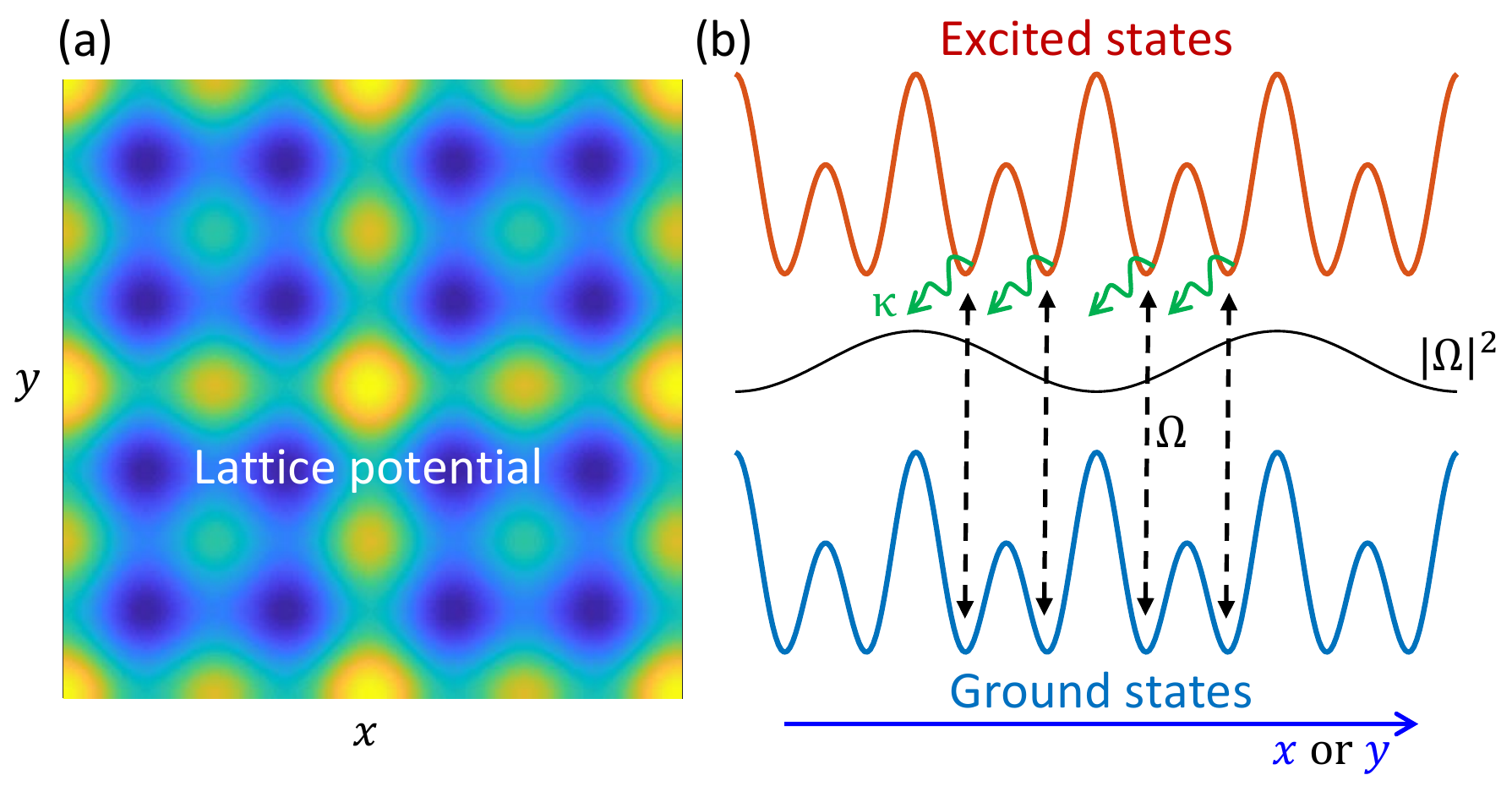}
\caption{(a) The optical lattice potential for the Hamiltonian Eq.~(1) in
the main text. (b) Scheme for engineering the staggered on-site particle
losses.}
\label{FigS:exp}
\end{figure}

\subsection{Experimental implementation}

In the main text, we have shown that the Hamiltonian Eq.~(1) in the main
text can be realized using coupled arrays of micro-ring cavities, as shown
in Fig.~1 in the main text. Each site is represented by a main cavity, which
is coupled to its neighbor cavities through the auxiliary coupling cavities
with controllable coupling strength and phase~\cite{arXiv1812.09304S}. The
loss/gain of each cavity can also be controlled independently~\cite%
{Zhao2018topologicalS}.

The Hamiltonian can also be realized using cold atoms in optical lattices,
with lattice potential shown in Fig.~\ref{FigS:exp} (a). The Hermitian part
can be realized within current techniques as proposed in Ref.~\cite%
{Benalcazar2017QuantizedS}. To obtain the on-site loss, we introduce the
resonance couplings between the ground state and the excited state with a
strong loss rate $\kappa $~\cite{arXiv1608.05061S,Ashida2017ParityS},
where the excited state feels the same lattice potential as the ground state
[see Fig.~\ref{FigS:exp} (b)]. The coupling $\Omega (x,y)$ between the
ground state and excited state 
gives rise to the effective loss for the ground state $2\gamma =\frac{\Omega
^{2}(x,y)}{\kappa}$~\cite{PhysRevA.85.032111S,PhysRevX.8.031079S}, and the
staggered loss can be controlled easily by $\Omega (x,y)$ [Fig.~\ref%
{FigS:exp} (b)]. Notice that the staggered loss configuration (without gain)
is equivalent to the staggered gain-loss configuration up to a constant.

We also would like to point out that both the coupled cavities and optical
lattices are able to realize the chiral non-Hermitian model with asymmetric
tunnelings. The optical-lattice scheme has been proposed in~\cite%
{arXiv1810.04067S}. For the coupled cavities, the asymmetric coupling can be
realized by introducing gain and loss to the two arms of the coupling
cavity, respectively~\cite{Sci.Rep.5.13376S}.

\end{widetext}

\end{document}